\documentclass[paper,longauth]{aa}

\usepackage{hyperref}
\hypersetup{
    bookmarks=true,                                                
    unicode=false,                                                 
    pdftoolbar=true,                                               
    pdfmenubar=true,                                               
    pdffitwindow=true,                                             
    pdftitle={The warm ionized gas in CALIFA early-type galaxies}, 
    pdfauthor={Author},                                            
    pdfsubject={Subject},                                          
    pdfnewwindow=true,                                             
    pdfkeywords={keywords},                                        
    colorlinks=true,                                               
    linkcolor=blue,                                                
    citecolor=blue,                                                
    filecolor=blue,                                                
    urlcolor=blue                                                  
}

\usepackage{amsmath}
\usepackage{natbib}
\usepackage{lscape}
\usepackage[section]{placeins}
\usepackage{wasysym}
\usepackage{graphics}
\usepackage{graphicx}
\usepackage{txfonts}
\usepackage{color}
\usepackage{times}
\usepackage{pgfkeys}
\usepackage{pgffor}

\definecolor{grey1}{rgb}{0.5,0.5,0.5}
\definecolor{greyJMG}{rgb}{0.23,0.23,0.23}
\definecolor{blueJMG}{rgb}{0.431372549,0.7568627451,0.9725490196}
\definecolor{orange}{rgb}{1.0,0.63,0}

\usepackage[T1]{fontenc}
\newcommand{\changefont}[3]{\fontfamily{#1}\fontseries{#2}\fontshape{#3}\selectfont}
\newfont{\vcap}{cmssdc10 scaled 1000}
\newfont{\lcap}{cmssdc10 scaled 1100}
\newfont{\nlx}{cmssdc10 scaled 900}
\newfont{\xnlx}{cmssdc10 scaled 800}
\newfont{\yx}{cmssdc10 scaled 700}
\newfont{\hvss}{cmssdc10 scaled 1540}
\newcommand{\rem}[1]{\nlx {#1}\normalfont}
\newcommand{\trem}[1]{\xnlx {#1}\normalfont}

\def\rr{{\sl R}$^{\star}$}
\def\kato{\rule[-1.25ex]{0cm}{1.25ex}}
\def\pano{\rule[0.0ex]{0cm}{2.5ex}}
\def\eqan{\begin{equation}}
\def\eqen{\end{equation}}

\def\h1{\ion{H}{i}}
\def\h2{\ion{H}{ii}}

\def\rr{{\sl R}$^{\star}$}

\def\ha{H$\alpha$}
\def\ewha{EW(\ha)}

\def\hb{H$\beta$}

\def\o5007{[O {\sc iii}] $\lambda$5007}
\def\ox{[O {\sc iii}]${\scriptstyle 5007}$}

\def\n2ha{[N\,{\sc ii}]/H$\alpha$}
\def\ln2ha{$\log$([N\,{\sc ii}]/H$\alpha$)}
\def\tn2ha{[N\,{\sc ii}]${\scriptstyle 6584}$/H$\alpha$}
\def\tln2ha{$\log$([N\,{\sc ii}]${\scriptstyle 6584}$/H$\alpha$)}
\def\o3hb{[O\,{\sc iii}\,]/H$\beta$}
\def\lo3hb{$\log$([O\,{\sc iii}]/H$\beta$)}
\def\to3hb{[O\,{\sc iii}]${\scriptstyle 5007}$/H$\beta$}
\def\tlo3hb{$\log$([O\,{\sc iii}]${\scriptstyle 5007}$/H$\beta$)}
\def\tlsiiha{$\log$($\Sigma$[S{\sc ii}]${\scriptstyle 6717,6731}$/H$\alpha$)}

\def\lsiiha{$\log$($\Sigma$[S{\sc ii}]/H$\alpha$)}

\def\e16{$10^{-16}~{\rm erg\,s^{-1}\,cm^{-2}}$}
\def\e17{$10^{-17}~{\rm erg\,s^{-1}\,cm^{-2}}$}
\def\tauha{$\tau$}
\def\tauha_ext{$\tau$}
\def\rPetr{$r_{\mathrm{p}}$}
\def\pAGBmin{$\mathrm{EW_{\star}^-}$}
\def\pAGBmax{$\mathrm{EW_{\star}^+}$}

\def\lyc{{{\sl Ly}$_{\rm c}$}}

\def\wism{\changefont{cmtt}{m}{it}wim\rm}
\def\wim{{\changefont{cmtt}{m}{it}{wim}}}

\def\plf{\changefont{cmtt}{m}{it}plf\rm}

\def\ha{H$\alpha$}
\def\ewha{EW(\ha)}

\def\tauha{$\tau$}
\def\tauha_ext{$\tau$}
\def\rPetr{$r_{\mathrm{p}}$}
\def\pAGBmin{{\sc ew}$_{\star}^-$}
\def\pAGBmax{{\sc ew}$_{\star}^+$}
\def\pAGBmean{$\langle \textrm{\sc ew}_{\star} \rangle$}
\def\sisp{\nlx sisp\rm} 
\def\isan{\nlx isan\rm} 
\def\mstar{${\cal M}_{\star}$}

\def\lfrac{${\cal L}_{\rm 5\,Gyr}$}
\def\porto3d{\sc Porto3D\rm}
\def\p3d{\sc Porto3D\rm}
\def\P3D{\sc Porto3D\rm}
\def\SL{{\sc Starlight}\rm}
\def\dvstars{$\Delta v_{\star}$}
\def\dvgas{$\Delta v_{\rm g}$}
\def\sisp{\nlx sisp\rm} 
 
\def\eqan{\begin{equation}}
\def\eqen{\end{equation}}

\def\h1{\ion{H}{i}}
\def\h2{\ion{H}{ii}}

\newcommand{\kmsec}{km~s$^{-1}$}
\newcommand{\msun}{M$_\odot$}
\newcommand{\zsun}{$Z_\odot$}

\newcommand{\sbb}{mag/$\sq\arcsec$}

\def\rr{{\sl R}$^{\star}$}
\def\P25{{\sl R}$_{\rm SF}$}
\def\E25{{\sl R}$_{\rm host}$}

\def\mstar{${\cal M}_{\star}$}

\def\tmass{$\langle t_{\star} \rangle_{{\cal M}}$}

\def\m5{${\cal M}_{\star,{\rm 5\,Gyr}}$}

\def\D4000{$D_{4000}$}
\def\ergsec{erg\,s$^{-1}$}
\def\uflux{erg\,s$^{-1}$\,cm$^{-2}$}
\def\ha{H$\alpha$}
\def\ewha{EW(\ha)}

\def\hb{H$\beta$}

\def\e16{$10^{-16}~{\rm erg\,s^{-1}\,cm^{-2}}$}
\def\e17{$10^{-17}~{\rm erg\,s^{-1}\,cm^{-2}}$}

\newcommand{\PutLabel}[3]{\put(#1,#2){#3}}

\pgfkeys{
  /plotETG/.is family, /plotETG,
  arqO/.estore in = \plotETGarqO,
  arqT/.estore in = \plotETGarqT,
  BPTs/.estore in = \plotETGBPTs,
  name/.estore in = \plotETGname,
  type/.estore in = \plotETGtype,
  morp/.estore in = \plotETGmorp,
  tauR/.estore in = \plotETGtauR,
  dist/.estore in = \plotETGdist,
  Magr/.estore in = \plotETGMagr,
  lumH/.estore in = \plotETGlumH,
  Mass/.estore in = \plotETGMass,
  desX/.estore in = \plotETGdesX,
  desY/.estore in = \plotETGdesY,
}
\newcommand\plotETG[2][1]{%
  \pgfkeys{/plotETG, #1}%

  \newpage
  \begin{figure*}
    \begin{picture}(18.4,27.0)

      \put(00.10,17.90){\includegraphics[width=0.270\textwidth, viewport=20 30 520 290]{Fig/\plotETGarqT_logHa_map.png}}
      \put(00.10,13.00){\includegraphics[width=0.270\textwidth, viewport=20 30 520 290]{Fig/\plotETGarqT_EWHa_map.png}}
      \put(00.10,08.10){\includegraphics[width=0.270\textwidth, viewport=20 30 520 290]{Fig/\plotETGarqT_vstar_map.png}}
      \put(00.10,03.20){\includegraphics[width=0.270\textwidth, viewport=20 30 520 290]{Fig/\plotETGarqT_vHaNII_map.png}}
      \put(-0.72,22.40){\includegraphics[width=0.270\textwidth, viewport=20 30 520 290]{Fig/\plotETGarqT_cont_map.png}}

      \put(06.90,22.80){\includegraphics[width=0.270\textwidth, viewport=20 30 520 290]{Fig/\plotETGarqT_pAGB_map.png}}
      \put(06.69,12.36){\includegraphics[trim=19.60cm 2.0cm -1.42cm 09.0cm, clip=true, width=6.25cm,height=6.16cm]{Fig/\plotETGarqO_R_vs_BPT_and_tau.pdf}}
      \put(06.20,12.70){\includegraphics[trim=02.30cm 3.6cm 15.1cm 10.0cm, clip=true, width=6.08cm,height=4.95cm]{Fig/\plotETGarqO_R_vs_I_and_EWHa.pdf}}

      \put(06.69,17.25){\includegraphics[trim=19.60cm 2cm -1.42cm 9.00cm, clip=true, width=6.25cm,height=6.16cm]{Fig/\plotETGarqO_R_vs_BPT_and_tau.pdf}}
      \put(06.20,17.50){\includegraphics[trim=02.30cm 11.4cm 15.1cm 02.2cm, clip=true, width=6.08cm,height=4.95cm]{Fig/\plotETGarqO_R_vs_I_and_EWHa.pdf}}

      \put(06.58,07.80){\includegraphics[trim=01.60cm 2.6cm 18.5cm 11.0cm, clip=true, width=5.00cm,height=4.62cm]{Fig/\plotETGarqO_bpt3a.pdf}}
      \put(06.58,02.90){\includegraphics[trim=10.45cm 2.6cm 09.0cm 11.0cm, clip=true, width=5.4cm,height=4.62cm]{Fig/\plotETGarqO_bpt3a.pdf}}

      \put(12.20,22.15){\includegraphics[trim=01.50cm 2.0cm 18.5cm 11.0cm, clip=true, width=5.03cm,height=5.00cm]{Fig/\plotETGarqO_R_vs_BPT_and_tau.pdf}}
      \put(12.02,17.25){\includegraphics[trim=10.10cm 2.0cm 09.5cm 11.0cm, clip=true, width=5.28cm,height=5.00cm]{Fig/\plotETGarqO_R_vs_BPT_and_tau.pdf}}
      \put(12.20,12.34){\includegraphics[trim=19.20cm 2.0cm -1.0cm 09.0cm, clip=true, width=6.10cm,height=6.19cm]{Fig/\plotETGarqO_R_vs_BPT_and_tau.pdf}}
      \put(12.60,03.14){\includegraphics[width=0.272\textwidth, viewport=20 30 520 290]{Fig/\plotETGarqT_L5Gyr_map.png}}       

      \PutLabel{00.20}{26.3}{\hvss a)}
      \PutLabel{00.20}{21.4}{\hvss b)}
      \PutLabel{00.20}{16.5}{\hvss c)}
      \PutLabel{00.20}{11.6}{\hvss d)}
      \PutLabel{00.20}{06.7}{\hvss e)}
      
      \PutLabel{07.00}{26.3}{\hvss f)}
      \PutLabel{10.60}{21.4}{\hvss g)}
      \PutLabel{10.60}{16.5}{\hvss h)}
      \PutLabel{07.00}{11.6}{\hvss i)}
      \PutLabel{07.00}{06.7}{\hvss j)}
      
      \PutLabel{12.60}{26.3}{\hvss k)}
      \PutLabel{12.60}{21.4}{\hvss l)}
      \PutLabel{12.60}{16.5}{\hvss m)}
      \PutLabel{12.60}{06.7}{\hvss n)}
      
      \PutLabel{12.50}{11.2}{\hvss \plotETGname}
      \PutLabel{12.50}{10.5}{\large Morphology NED: \plotETGmorp}
      \PutLabel{12.50}{10.0}{\large ETG type~\plotETGtype}
      \PutLabel{12.50}{09.5}{\large $\langle \tau$(H$\alpha$)$_{\rm e} \rangle$ = \plotETGtauR}
      \PutLabel{12.50}{09.0}{\large $D$ = \plotETGdist\ Mpc | $M_{r}$ = \plotETGMagr\ mag}
      \PutLabel{12.50}{08.5}{\large L(\ha) = \plotETGlumH\ $\times 10^{39}~{\rm erg\,s^{-1}}$}
      \PutLabel{12.50}{08.0}{\large log \mstar = \plotETGMass\ [M$_\odot$]}

      \PutLabel{00.00}{27.05}{\color{greyJMG}\yx FoV 78\arcsec $\times$ 72\arcsec\color{black}}
      \PutLabel{+6.35}{23.45}{\rotatebox{90}{\color{greyJMG}\yx EW(H$\alpha$) map into 3 intervals\color{black}}}

      \PutLabel{+6.8}{27.05}{\bf \color{greyJMG}\yx EW(H$\alpha$) \color{black}}
      \PutLabel{+8.7}{27.05}{\bf \color{greyJMG}\yx
        \textcolor{blue}{\bf \yx $\leq$0.5} / \textcolor{blueJMG}{\bf \yx
          0.5--2.4} / \textcolor{orange}{\bf \yx $>$2.4} {\tiny [\text{\AA}]} \color{black}}
      
      \PutLabel{+5.30}{22.63}{\rotatebox{90}{\color{greyJMG}\yx log(stellar continuum [10$^{{\yx -16}}$ \uflux])\color{black}}}
      \PutLabel{+5.30}{18.40}{\rotatebox{90}{\color{greyJMG}\yx H$\alpha$ flux [10$^{\yx -16}$ \uflux]\color{black}}}      
      \PutLabel{+5.30}{14.30}{\rotatebox{90}{\color{greyJMG}\yx EW(H$\alpha$) [$\AA$]\color{black}}}
      \PutLabel{+5.42}{08.40}{\rotatebox{90}{\color{greyJMG}\yx stellar line-of-sight velocity [km/s]\color{black}}}      
      \PutLabel{+5.42}{03.30}{\rotatebox{90}{\color{greyJMG}\yx H$\alpha$+[N{\sc ii}] line-of-sight velocity [km/s]\color{black}}}

      \PutLabel{-0.40}{08.20}{\rotatebox{90}{\color{blue}\yx Photometric minor axis (blue lines)\color{black}}}      

      \PutLabel{+5.90}{19.20}{\rotatebox{90}{\yx H$\alpha$ log (I/I$_0$)}}      
      \PutLabel{+5.90}{14.30}{\rotatebox{90}{\yx EW(H$\alpha$) [$\AA$]}}      
      \PutLabel{+8.30}{12.35}{\yx Radius (arcsec)} 

      \PutLabel{+6.02}{9.30}{\rotatebox{90}{\yx log [OIII]5007/H$\beta$}}      
      \PutLabel{+8.28}{07.50}{\yx log [NII]6584/H$\alpha$} 

      \PutLabel{+6.02}{4.30}{\rotatebox{90}{\yx log [OIII]5007/H$\beta$}}      
      \PutLabel{+7.90}{02.50}{\yx log [$\Sigma$[SII]6717,6731/H$\alpha$} 

      \PutLabel{+14.00}{12.35}{\yx Radius (arcsec)} 
      \PutLabel{+11.70}{23.90}{\rotatebox{90}{\yx log [OIII]5007/H$\beta$}}      
      \PutLabel{+11.70}{18.90}{\rotatebox{90}{\yx log [NII]6584/H$\alpha$}}      
      \PutLabel{+11.70}{14.72}{\rotatebox{90}{\yx log $\tau$}}      

      \PutLabel{+18.0}{02.95}{\rotatebox{90}{\color{greyJMG}\yx luminosity fraction of stars with t$\leq$5 Gyr [\%]\color{black}}}
    \end{picture}

    \vspace*{-25mm}
    \caption[]{\small \textbf{\plotETGname}: 2D maps (panels
      \rem{a}-\rem{f} and \rem{n}), radial intensity and EW of \ha\ (panels
      \rem{g} and \rem{h}), BPT diagrams (panels \rem{i} and \rem{j}),
      radial distribution of diagnostic line ratios (\rem{k} and \rem{l}),
      and the $\tau$ ratio (panel \rem{m}). In all 2D maps north is up and
      east to the left. The horizontal bar corresponds to
      20\arcsec. }
    \label{fig:\plotETGBPTs}
  \end{figure*}

  \clearpage  
}

\pgfkeys{
  /plotCom/.is family, /plotCom,
  arqO/.estore in = \plotComarqO,
  arqT/.estore in = \plotComarqT,
  BPTs/.estore in = \plotComBPTs,
  name/.estore in = \plotComname,
  morp/.estore in = \plotCommorp,
  tauR/.estore in = \plotComtauR,
  dist/.estore in = \plotComdist,
  Magr/.estore in = \plotComMagr,
  lumH/.estore in = \plotComlumH,
  Mass/.estore in = \plotComMass,
}
\newcommand\plotCom[2][1]{%
  \pgfkeys{/plotCom, #1}%

  \newpage
  \begin{figure*}
    \begin{picture}(18.4,27.0)
      \put(00.10,17.90){\includegraphics[width=0.270\textwidth, viewport=20 30 520 290]{Fig/\plotComarqT_logHa_map.png}}
      \put(00.10,13.00){\includegraphics[width=0.270\textwidth, viewport=20 30 520 290]{Fig/\plotComarqT_EWHa_map.png}}
      \put(00.10,08.10){\includegraphics[width=0.270\textwidth, viewport=20 30 520 290]{Fig/\plotComarqT_vstar_map.png}}
      \put(00.10,03.20){\includegraphics[width=0.270\textwidth, viewport=20 30 520 290]{Fig/\plotComarqT_vHaNII_map.png}}
      \put(-0.72,22.40){\includegraphics[width=0.270\textwidth, viewport=20 30 520 290]{Fig/\plotComarqT_cont_map.png}}

      \put(06.90,22.80){\includegraphics[width=0.270\textwidth, viewport=20 30 520 290]{Fig/\plotComarqT_pAGB_map.png}}
      \put(06.69,12.36){\includegraphics[trim=19.60cm 2.0cm -1.42cm 09.0cm, clip=true, width=6.25cm,height=6.16cm]{Fig/\plotComarqO_R_vs_BPT_and_tau.pdf}}
      \put(06.20,12.70){\includegraphics[trim=02.30cm 3.6cm 15.1cm 10.0cm, clip=true, width=6.08cm,height=4.95cm]{Fig/\plotComarqO_R_vs_I_and_EWHa.pdf}}

      \put(06.69,17.25){\includegraphics[trim=19.60cm 2cm -1.42cm 9.00cm, clip=true, width=6.25cm,height=6.16cm]{Fig/\plotComarqO_R_vs_BPT_and_tau.pdf}}
      \put(06.20,17.50){\includegraphics[trim=02.30cm 11.4cm 15.1cm 02.2cm, clip=true, width=6.08cm,height=4.95cm]{Fig/\plotComarqO_R_vs_I_and_EWHa.pdf}}

      \put(06.58,07.80){\includegraphics[trim=01.60cm 2.6cm 18.5cm 11.0cm, clip=true, width=5.00cm,height=4.62cm]{Fig/\plotComarqO_bpt3a.pdf}}
      \put(06.58,02.90){\includegraphics[trim=10.45cm 2.6cm 09.0cm 11.0cm, clip=true, width=5.4cm,height=4.62cm]{Fig/\plotComarqO_bpt3a.pdf}}

      \put(12.20,22.15){\includegraphics[trim=01.50cm 2.0cm 18.5cm 11.0cm, clip=true, width=5.03cm,height=5.00cm]{Fig/\plotComarqO_R_vs_BPT_and_tau.pdf}}
      \put(12.02,17.25){\includegraphics[trim=10.10cm 2.0cm 09.5cm 11.0cm, clip=true, width=5.28cm,height=5.00cm]{Fig/\plotComarqO_R_vs_BPT_and_tau.pdf}}
      \put(12.20,12.34){\includegraphics[trim=19.20cm 2.0cm -1.0cm 09.0cm, clip=true, width=6.10cm,height=6.19cm]{Fig/\plotComarqO_R_vs_BPT_and_tau.pdf}}
      \put(12.60,03.14){\includegraphics[width=0.272\textwidth, viewport=20 30 520 290]{Fig/\plotComarqT_L5Gyr_map.png}}       

      \PutLabel{00.20}{26.3}{\hvss a)}
      \PutLabel{00.20}{21.4}{\hvss b)}
      \PutLabel{00.20}{16.5}{\hvss c)}
      \PutLabel{00.20}{11.6}{\hvss d)}
      \PutLabel{00.20}{06.7}{\hvss e)}
      
      \PutLabel{07.00}{26.3}{\hvss f)}
      \PutLabel{10.60}{21.4}{\hvss g)}
      \PutLabel{10.60}{16.5}{\hvss h)}
      \PutLabel{07.00}{11.6}{\hvss i)}
      \PutLabel{07.00}{06.7}{\hvss j)}
      
      \PutLabel{12.60}{26.3}{\hvss k)}
      \PutLabel{12.60}{21.4}{\hvss l)}
      \PutLabel{12.60}{16.5}{\hvss m)}
      \PutLabel{12.60}{06.7}{\hvss n)}
      
      \PutLabel{12.50}{11.2}{\hvss \plotComname}
      \PutLabel{12.50}{10.5}{\large \textcolor{red}{\bf Comparison Sample}}
      \PutLabel{12.50}{10.0}{\large Morphology NED: \plotCommorp}
      \PutLabel{12.50}{09.5}{\large $\langle \tau$(H$\alpha$)$_{\rm e} \rangle$ = \plotComtauR}
      \PutLabel{12.50}{09.0}{\large $D$ = \plotComdist\ Mpc | $M_{r}$ = \plotComMagr\ mag}
      \PutLabel{12.50}{08.5}{\large L(\ha) = \plotComlumH\ $\times 10^{39}~{\rm erg\,s^{-1}}$}
      \PutLabel{12.50}{08.0}{\large log \mstar = \plotComMass\ [M$_\odot$]}

      \PutLabel{00.00}{27.05}{\color{greyJMG}\yx FoV 78\arcsec $\times$ 72\arcsec\color{black}}
      \PutLabel{+6.35}{23.45}{\rotatebox{90}{\color{greyJMG}\yx EW(H$\alpha$) map into 3 intervals\color{black}}}

      \PutLabel{+6.8}{27.05}{\bf \color{greyJMG}\yx EW(H$\alpha$) \color{black}}
      \PutLabel{+8.7}{27.05}{\bf \color{greyJMG}\yx
        \textcolor{blue}{\bf \yx $\leq$0.5} / \textcolor{blueJMG}{\bf \yx
          0.5--2.4} / \textcolor{orange}{\bf \yx $>$2.4} {\tiny [\text{\AA}]} \color{black}}

      \PutLabel{+5.30}{22.63}{\rotatebox{90}{\color{greyJMG}\yx log(stellar continuum [10$^{{\yx -16}}$ \uflux])\color{black}}}
      \PutLabel{+5.30}{18.40}{\rotatebox{90}{\color{greyJMG}\yx H$\alpha$ flux [10$^{\yx -16}$ \uflux]\color{black}}}      
      \PutLabel{+5.30}{14.30}{\rotatebox{90}{\color{greyJMG}\yx EW(H$\alpha$) [$\AA$]\color{black}}}
      \PutLabel{+5.42}{08.40}{\rotatebox{90}{\color{greyJMG}\yx stellar line-of-sight velocity [km/s]\color{black}}}      
      \PutLabel{+5.42}{03.30}{\rotatebox{90}{\color{greyJMG}\yx H$\alpha$+[N{\sc ii}] line-of-sight velocity [km/s]\color{black}}}

      \PutLabel{+5.90}{19.20}{\rotatebox{90}{\yx H$\alpha$ log (I/I$_0$)}}      
      \PutLabel{+5.90}{14.30}{\rotatebox{90}{\yx EW(H$\alpha$) [$\AA$]}}      
      \PutLabel{+8.30}{12.35}{\yx Radius (arcsec)} 

      \PutLabel{+6.02}{9.30}{\rotatebox{90}{\yx log [OIII]5007/H$\beta$}}      
      \PutLabel{+8.28}{07.50}{\yx log [NII]6584/H$\alpha$} 

      \PutLabel{+6.02}{4.30}{\rotatebox{90}{\yx log [OIII]5007/H$\beta$}}      
      \PutLabel{+7.90}{02.50}{\yx log [$\Sigma$[SII]6717,6731/H$\alpha$} 

      \PutLabel{+14.00}{12.35}{\yx Radius (arcsec)} 
      \PutLabel{+11.70}{23.90}{\rotatebox{90}{\yx log [OIII]5007/H$\beta$}}      
      \PutLabel{+11.70}{18.90}{\rotatebox{90}{\yx log [NII]6584/H$\alpha$}}      
      \PutLabel{+11.70}{14.72}{\rotatebox{90}{\yx log $\tau$}}      

      \PutLabel{+18.0}{02.95}{\rotatebox{90}{\color{greyJMG}\yx luminosity fraction of stars with t$\leq$5 Gyr [\%]\color{black}}}

      \PutLabel{-0.40}{08.20}{\rotatebox{90}{\color{blue}\yx Photometric minor axis (blue lines)\color{black}}}      

    \end{picture}

    \vspace*{-25mm}
    \caption[]{\small \textbf{\plotComname}: 2D maps (panels
      \rem{a}-\rem{f} and \rem{n}), radial intensity and EW of \ha\ (panels
      \rem{g} and \rem{h}), BPT diagrams (panels \rem{i} and \rem{j}),
      radial distribution of diagnostic line ratios (\rem{k} and \rem{l}),
      and the $\tau$ ratio (panel \rem{m}). In all 2D maps north is up and
      east to the left. The horizontal bar corresponds to 20\arcsec. }
    \label{fig:\plotComBPTs}
  \end{figure*}

  \clearpage  
}

\newcounter{qub}
\begin{document}

\title{The warm ionized gas in CALIFA early-type galaxies}
\subtitle{2D emission-line patterns and kinematics for 32 galaxies}

\author{
J.M. Gomes\inst{\ref{inst1}} 
\and 
P.~Papaderos\inst{\ref{inst1}}
\and 
C.~Kehrig\inst{\ref{inst2}}
\and
J.~M.~V\'{\i}lchez\inst{\ref{inst2}}
\and
M.~D.~Lehnert\inst{\ref{inst3}}
\and
S.~F.~S\'anchez\inst{\ref{inst2},\ref{inst4}}
\and 
B.~Ziegler\inst{\ref{inst5}}
\and 
I.~Breda\inst{\ref{inst1}}
\and 
S.N.~dos Reis\inst{\ref{inst1}}
\and 
J.~Iglesias-P\'aramo\inst{\ref{inst2},\ref{inst6}}
\and 
J.~Bland-Hawthorn\inst{\ref{inst7}}
\and 
L.~Galbany\inst{\ref{inst8},\ref{inst9}}
\and
D.~J.~Bomans\inst{\ref{inst10},\ref{inst11}}
\and
F.~F.~Rosales-Ortega\inst{\ref{inst12}}
\and
R.~Cid Fernandes\inst{\ref{inst13}}
\and 
C.~J.~Walcher\inst{\ref{inst14}}
\and
J.~Falc\'on-Barroso\inst{\ref{IAC},\ref{UnivLaguna}}
\and
R.~Garc\'ia-Benito\inst{\ref{inst2}}
\and
I.~M\'arquez\inst{\ref{inst2}}
\and
A.~del Olmo\inst{\ref{inst2}}
\and
J.~Masegosa\inst{\ref{inst2}}
\and 
M.~Moll\'a\inst{\ref{inst15}}
\and
R.~A.~Marino\inst{\ref{inst16},\ref{inst17}}
\and
R.~M.~Gonz\'alez Delgado\inst{\ref{inst2}}
\and
\'A.~R.~L\'opez-S\'anchez\inst{\ref{inst18},\ref{inst19}}
\and
the CALIFA collaboration
}
\offprints{Jean Michel Gomes; jean@astro.up.pt}
\institute{
Instituto de Astro{f\'\i}sica e Ci{\^e}ncias do Espa\c{c}o, Universidade do Porto,
Centro de Astrof{\'\i}sica da Universidade do Porto, Rua das Estrelas, 4150-762 Porto,
Portugal\label{inst1}
\and
Instituto de Astrof{\'\i}sica de Andaluc{\'\i}a (CSIC), Glorieta de la
Astronom\'{\i}a s/n Aptdo. 3004, E18080-Granada, Spain\label{inst2}
\and
Institut d'Astrophysique de Paris, UMR 7095, CNRS, Universit\'{e} Pierre et Marie Curie, 98 bis boulevard Arago, 75014 Paris, France\label{inst3}
\and
Centro Astron\'omico Hispano Alem\'an de Calar Alto (CSIC-MPG), C/ Jes\'us Durb\'an Rem\'on 2-2, E-4004 Almer\'ia, Spain\label{inst4}
\and
University of Vienna, T\"{u}rkenschanzstrasse 17, 1180 Vienna, Austria\label{inst5}
\and
Estaci\'on Experimental de Zonas Aridas (CSIC), Ctra. de Sacramento s.n., La
Ca\~nada, Almer\'ia, Spain\label{inst6}
\and
Sydney Institute for Astronomy, University of Sydney, NSW 2006,
Australia\label{inst7}
\and
Millennium Institute of Astrophysics, Chile\label{inst8}
\and
Departamento de Astronom\'ia, Universidad de Chile, Casilla 36-D, Santiago,
Chile\label{inst9}
\and
Astronomical Institute of the Ruhr-University Bochum\label{inst10}
\and
RUB Research Department Plasmas with Complex Interactions\label{inst11}
\and
Instituto Nacional de Astrof\'isica, \'Optica y Electr\'onica, Luis E. Erro 1, 72840 Tonantzintla, Puebla, Mexico\label{inst12}
\and
Departamento de F\'isica, Universidade Federal de Santa Catarina, PO Box 476, 88040-900, Florian\'opolis, SC, Brazil\label{inst13} 
\and
Leibniz-Institut f\"ur Astrophysik Potsdam (AIP), An der Sternwarte 16, D-14482 Potsdam, Germany\label{inst14}
\and
Instituto de Astrof\'isica de Canarias, V\'ia L\'actea s/n, La Laguna, Tenerife, Spain\label{IAC}
\and
Departamento de Astrof\'isica, Universidad de La Laguna, E-38205 La Laguna, Tenerife, Spain\label{UnivLaguna}
\and
CIEMAT, Avda. Complutense 40, 28040 Madrid, Spain\label{inst15}
\and
CEI Campus Moncloa, UCM-UPM, Departamento de Astrof\'isica y CC. de la Atm\'osfera, Facultad de CC. F\'isicas, Universidad Complutense de Madrid,
Avda. Complutense s/n, 28040 Madrid, Spain\label{inst16}
\and
Department of Physics, Institute for Astronomy, ETH Z\"urich, CH-8093 Z\"urich, Switzerland\label{inst17}
\and
Australian Astronomical Observatory, PO Box 915, North Ryde,
NSW 1670, Australia\label{inst18}
\and
Department of Physics and Astronomy, Macquarie University, NSW
2109, Australia\label{inst19}
}
\date{Received ?? ; Accepted ??}

\abstract
{ The morphological, spectroscopic and kinematical properties of the warm
  interstellar medium (\wim) in early-type galaxies (ETGs) hold key
  observational constraints to nuclear activity and the buildup history of
  these massive, quiescent systems.  High-quality integral field
  spectroscopy (IFS) data with a wide spectral and spatial coverage, such
  as those from the CALIFA survey, offer an unprecedented opportunity for
  advancing our understanding of the \wim\ in ETGs.  }
{ This article centers on a 2D investigation of the \wim\ component in 32
  nearby ($\la$150~Mpc) ETGs from CALIFA, complementing a previous 1D
  analysis of the same sample (Papaderos et al. 2013; P13).}
{ The analysis presented here includes \ha\ intensity and equivalent width
  (EW) maps and radial profiles, diagnostic emission-line ratios, besides
  ionized-gas and stellar kinematics. It is supplemented by $\tau$-ratio
  maps as an efficient means to quantify the role of photoionization by the
  post-AGB stellar component, as compared to alternative mechanisms (e.g.,
  AGN, low-level star formation).}
{ Confirming and strengthening our conclusions in P13, we find that ETGs
  span a broad continuous sequence in the properties of their \wim,
  exemplified by two characteristic classes. The first one (type~i)
  comprises systems with a nearly constant \ewha\ in their extranuclear
  component, in quantitative agreement with (even though, no proof for) the
  hypothesis of photoionization by the post-AGB stellar component being the
  main driver of extended \wim\ emission.  The second class (type~ii)
  stands for virtually \wim-evacuated ETGs with a very low ($\leq$0.5~\AA),
  outwardly increasing \ewha.  These two classes appear indistinguishable
  from one another by their LINER-specific emission-line ratios in their
  extranuclear component.  Here we extend the tentative classification we
  proposed previously by the type~i+, which is assigned to a subset of
  type~i ETGs exhibiting ongoing low-level star-forming activity in their
  periphery.  This finding along with faint traces of localized star
  formation in the extranuclear component of several of our sample galaxies
  points to a non-negligible contribution by OB stars to the global
  ionizing photon budget in ETGs.
  Additionally, our data further highlight the diversity of ETGs with
  respect to their gaseous and stellar kinematics.  Whereas, in one half of
  our sample, gas and stars show similar (yet not necessarily identical)
  velocity patterns, both dominated by rotation along the major galaxy
  axis, our analysis also documents several cases of kinematical decoupling
  between gas and stars, or rotation along the minor galaxy axis.
  We point out that the generally very low ($\la$1~\AA) \ewha\ of ETGs make
  necessary a careful quantitative assessment of potential observational
  and analysis biases in studies of their \wim.  Since, with standard
  emission-line fitting tools, Balmer emission lines become progressively
  difficult to detect below an \ewha$\sim$3~\AA, our current understanding
  of the presence, and 2D emission patterns and kinematics of the diffuse
  \wim\ ETGs may be severely incomplete.  We demonstrate that, at the
  typical emission-line detection threshold of $\sim$2~\AA\ in previous
  studies, most of the extranuclear \wim\ emission in an ETG may evade
  detection, which could in turn prompt its classification as an entirely
  gas-devoid system.}
{ This study adds further observational evidence for a considerable
  heterogeneity among ETGs with regard to the physical properties and 2D
  kinematics of their extended \wim\ component, and underscores the
  importance of IFS studies of these systems over their entire optical
  extent.}


\maketitle
\markboth {Gomes et al.}{The warm ionized gas in CALIFA early-type galaxies}

\section{Introduction \label{intro}}
The featureless appearance of early-type galaxies (ETGs) has for decades
sustained the view that these systems are largely `simple', both in terms
of their mass assembly history and kinematics, formed early on, and having
undergone little evolution over the past several Gyr
\citep[e.g.,][]{mat71,bre78,whi83}.  However, our understanding of the
buildup of these systems has undergone a substantial revision over the last
years, as ground-based and satellite multi-wavelength data have gradually
revealed a great deal of structural and kinematical diversity in
present-day ETGs \citep[e.g.,][]{Bender89,Kormendy09}.

For instance, careful photometric studies have revealed in many of these
systems a rich variety of low-surface brightness non-axis-symmetric
features, such as ripples and shells \citep[e.g.,][and references
  therein]{SchweizerSeitzer88,Struck1999}, pointing to a tumultuous
assembly history, with multiple minor merger episodes, some of which may
have provided the gas fuel for rejuvenating these `old and dead' galaxies
with low-level star formation (SF) \citep[see, e.g.,][and references
  therein]{Trager2000,Rampazzo07,Kaviraj08,Clemens09,Thomas10}.  Indeed,
the UV-to-optical colors of several ETGs are consistent with a prolonged
phase of low-level SF, in which 1\% to 3\% of the stellar mass
\mstar\ could have formed in the past one Gyr \citep{Kaviraj07}.

With the advent of integral field spectroscopy (IFS), spatially resolved
analyses have further highlighted the diversity and complexity of ETGs.
For instance, following the initial study of \object{NGC 3377} by
\cite{Bacon01}, the SAURON project has given a decisive contribution to the
systematization of the kinematical properties of ETGs, among others, by
uncovering kinematically decoupled or even counter-rotating cores in
several of these systems \citep[e.g.,][]{Sarzi06}.  Further insights into
the connection between kinematics and other ETG properties are being added
with the ATLAS$^{\rm 3D}$ IFS survey \citep{Cappellari11-Atlas3D},
supplementing previous work with information on, e.g., the relation between
the specific angular momentum and the shape of surface brightness profiles
\citep[e.g.,][]{Krajnovic12-Atlas3D}, or that between the degree of
rotational support and X-ray luminosity \citep{Sarzi12-Atlas3D-Xray}.

Another key insight has been the presence of faint nebular emission in the
majority of ETGs studied with longslit and single-aperture SDSS spectroscopy
\citep[e.g.,][]{Phillips1986,dem84,kim89,tri91,ani10,YanBlanton2012} or
narrow-band imaging \citep{Finkelmann2010,Kulkarni2014}. 
In particular, SAURON, has detected extended nebular emission in $\sim$75\% of the
studied ETGs (48 in total), strengthening previous evidence for the presence of an
ubiquitous warm (T$\sim$10$^4$ K) interstellar medium (\wism) component in
these systems, and adding new information on its kinematics. 
For example, \citet[][see also Sarzi et al. 2007]{Sarzi06} reported the detection of
extended Balmer H$\beta$ emission in ETGs down to an equivalent width (EW) of
$\simeq$0.1~\AA, and demonstrated a puzzling variety of morphologies and
kinematics, with striking cases of kinematical decoupling between gas and
stellar motions. 

Using IFS data from the {\sl Calar Alto Legacy Integral Field Area}
survey \citep[CALIFA,][]{Sanchez2012}, the presence of an extended
\wim\ component in ETGs was first demonstrated by \citet[][hereafter K12]{K12},
and subsequently by \cite{P13} and \cite{Singh13}.
CALIFA is the first IFS survey simultaneously permitting full coverage of the
optical spectral range and spatial extent of a representative sample of
Hubble-type galaxies in the local universe \citep[see][for description of the
sample selection]{Walcher14}. The wide spectral and spatial coverage of
CALIFA is a key advantage over SAURON and ATLAS$^{\rm 3D}$, which target a 
narrow spectral interval encompassing the \hb\ and \ox\ lines over a field
of view (FoV) of 33\arcsec$\times$41\arcsec ($\sim$1/3 of that of CALIFA). 
As recently pointed out by \citet{Arnold13} and \citet{G14a}, a full mapping of
the optical galaxy size is essential for an unbiased determination of the
properties of gas and stars in ETGs.

Spatially resolved IFS studies of emission-line EWs and line-flux ratios of
the \wim\ also yield key constraints on the energy balance and gas
excitation mechanisms in ETGs -- an issue being debated for at least the
last three decades.  Most ETG nuclei show faint nebular emission with an
\ha\ equivalent width of typically EW(\ha)$\leq$10 \AA\ \citep[see,
  e.g.,][]{ani10} and are classified as {\sl low-ionization nuclear
  emission-line regions}\footnote{Since it is now recognized that
  LINER-specific emission-line ratios are not confined to galaxy nuclei
  \citep[e.g.,][]{Sharp2010,K12,P13,Singh13}, a more adequate acronym would
  be {\sl low-ionization emission-line regions} (LIERs).}
\citep[LINERs;][]{Heckman1980} on the basis of optical diagnostic line
ratios after \citet[][referred to in the following as BPT ratios]{bpt81}
and \citep[][hereafter VO]{VO1987}.  One widely favored mechanism for the
\wim\ excitation in ETG nuclei involves `weak', {\sl low-luminosity active
  galactic nuclei} (LLAGN) \citep[cf, e.g.,][]{Ho1999}, powered by
radiatively inefficient (sub-Eddington) accretion onto a central
super-massive black hole (SMBH), a process that is thought to be governed
by the absence of the broad-line region and its surrounding torus, as in
classical Active Galactic Nuclei (AGN) \citep[][and references
  therein]{Ho2008}.
  
Other hypotheses involve fast shocks
\citep[e.g.,][]{DopitaSutherland1995,Allen2008}, photoionization by ongoing
low-level SF \citep[e.g.,][]{Schawinski2007,Shapiro2010} and hot evolved
($\geq 10^8$ yr) post-asymptotic giant branch (pAGB) stars
\citep[e.g.,][]{tri91,bin94,macc96,sta08,cid10,cid11,Sarzi10,YanBlanton2012}.
The latter mechanism has attracted a great deal of interest in the last
years and was even proposed to be the sole source of gas excitation in
ETG/LINERs.  Turning into this subject, following earlier work by
\cite{dSA90} and \cite{tri91}, \cite{bin94} have first studied
quantitatively with evolutionary synthesis models the ionizing output from
white dwarfs and hot pAGB stars\footnote{For the sake of brevity we will
  refer to in the following as `pAGB stars' to all ionizing point sources
  arising at and beyond the post-AGB phase.}, and demonstrated that these
sources are capable of producing the ionizing radiation for simultaneously
explaining the low nuclear EWs and LINER characteristics of ETG nuclei.
Several observational studies have added support to the pAGB
photoionization hypothesis. For example, \citet{Sarzi10}, using SAURON IFS
data, conclude that pAGB stars, instead of fast shocks, are the main source
of ionizing photons in ETGs.  This view was reinforced, at least as far
extranuclear \wism\ emission is concerned, by our study in K12, and, more
recently, by a combined study of Sloan Digital Sky Survey
\citep[SDSS;][]{yor00} and Palomar survey \citep{Ho1995} spectra by
\cite{YanBlanton2012}. Likewise, \cite{Eracleous2010} point out that pAGB
stars provide more ionizing photons than AGN in more than half of the
LINERs and can account for the observed line emission in 1/3 of these
systems.  On the other hand, arguments against the predominance of pAGB
photoionization in ETGs/LINERs were discussed in other studies.  For
instance, \citet{ani10} conclude from the analysis of long-slit spectra for
65 ETGs that in only $\sim$1/5 of their sample can the nuclear LINER
emission be attributed to pAGB photoionization alone, and excitation by a
low-accretion rate AGN or fast shocks is likely necessary for the majority
of ETG nuclei.  The pAGB component as the dominant source of
photoionization in ETG nuclei has also been disputed by, e.g.,
\citet{Ho2008}, on the basis of various arguments, one of them being that
line emission in these systems tends to be very centrally concentrated.

Following the seminal theoretical work by \citet{bin94}, the pAGB
photoionization hypothesis has been recently further investigated in e.g.,
\citet[][]{SodreStasinska1999,sta08,cid10,cid11,Stasinska15} and confronted
with SDSS spectroscopic data.  As \cite{sta08} argue, a significant
fraction of ETGs are in fact `retired' galaxies, i.e.  systems that stopped
forming stars, and whose ionizing field is powered solely by pAGB
sources. According to the classification by \cite{cid11}, this large
population of retired ETG/LINERs is characterized by a \n2ha\ ratio of
$\ga$1 (similar to classical AGN) yet faint nebular emission (0.5 $\ga$
EW(\ha)~[\AA] $\la$ 3.0). Quite importantly, the generally satisfactorily
agreement between pAGB photoionization models and single-aperture SDSS data
has called into question the necessity of any additional energy source (SF,
AGN) in ETGs, and led even to the rejection of `weak' (low-accretion rate,
low-luminosity) AGN in these systems: As emphatically pointed out in
\citet{cid11}, retired galaxies (i.e. low-EW(\ha) ETG/LINERs) are actually
fake AGN, {\it erroneously counted as AGN, and leading to the illusion of a
  Seyfert-LINER dichotomy in the AGN population}.

In the framework of the CALIFA collaboration, two recent studies
\citep[P13, ][]{Singh13} have confirmed and strengthened the conclusion by
K12 that LINER-like nebular emission in ETGs typically extends several kpc
away from the galaxy nuclei. They also lent further support, on the basis
of distinctively different arguments, to the idea that pAGB photoionization
can be an important ingredient of the \wim\ excitation on galactic scales.
More specifically, P13 studied the radial distribution of the \ewha\ over
nearly the entire optical extent of 32 ETGs and demonstrated that this
quantity shows a nearly constant value of $\sim$1~\AA\ in the extranuclear
component of $\sim$40\% of their sample (type~i ETGs in their
classification), in quantitative agreement with the diffuse floor of
ionization expected from the pAGB background.  A second test made involves
the $\tau$ ratio \citep[K12, see also][for similar
  definitions]{bin94,cid11}, defined as the inverse ratio of the observed
\ha\ luminosity to that predicted from pAGB photoionization when assuming
standard conditions in the gas and case~B recombination.  The nearly
constant $\tau\simeq 1$ determined in their extranuclear component of
type~i ETGs implies that the Lyman continuum photon rate from pAGB stars is
energetically capable of sustaining the observed low-level \wim\ emission.

On the other hand, P13 have shown that ETGs in their majority display more
complex 2D gas emissivity patterns than what is expected from in situ
(case~B) recombination of the \lyc\ output from the pAGB component.
Specifically, in $\sim$60\% of the ETGs in their sample (type~ii in their
classification) the $\tau$ ratio was determined to be between $\geq 2$ and
up to $\ga$50, implying that the bulk of \lyc\ photons produced by pAGB
stars escapes without being locally reprocessed into nebular emission.
They interpreted this and the positive radial \ewha\ gradients of type~ii
ETGs as the result of the interplay between various physical parameters
determining the 3D characteristics of the warm and hot gas component in
ETGs, such as, e.g., porosity, filling factor and ionization parameter.
The fact that type~i and type~ii ETGs are almost indistinguishable from one
another by their LINER characteristics, despite a difference of $\geq$1~dex
in their $\tau$ (hence, also their 3D gas structure and \lyc\ photon escape
fraction) has led P13 to conjecture that, in the 3D geometry of ETGs,
classical BPT diagnostics become degenerate over the area of the diagrams
which cover galaxies with LINER-like ratios. Additionally, as first pointed
out in P13, a far-reaching implication of the large \lyc\ escape fraction
in type~ii ETGs is that these systems could host significant
accretion-powered nuclear activity that evades detection through optical
spectroscopy. This is because, in the presence of extensive \lyc\ escape,
the luminosities of nebular emission lines commonly used as diagnostics of
accretion-powered nuclear activity and/or to which SMBHs mass
determinations are tied, are reduced by at least one order of
magnitude. For the same reason, a low nuclear EW(\ha) (0.5\dots3 \AA) is
not per se compelling evidence for the \wim\ emission in ETGs being solely
powered by pAGB stars, even though their ionizing photon output is able to
sustain their diffuse \ha\ emission.  As we recently pointed out in
\citet{G14b}, it is imaginable that the combined effect of different gas
excitation sources (SF, pAGB, AGN and shocks) with geometry-dependent
\lyc\ leakage can lead to a nearly constant \ewha$\sim$1~\AA, mimicking a
predominance of pAGB photoionization all over the extranuclear component of
type~i ETGs.

Large-FoV, wide spectral coverage IFS data, as those from CALIFA, obviously
offer a tremendous potential for further elucidating the nature and
excitation mechanisms of the \wism\ in ETGs.  In this article, we discuss
in more detail our methodology in P13 and extend our analysis by a concise
summary of the 2D properties of the nebular and stellar component in our
ETG sample, as derived with our IFS data processing pipeline \p3d.  This
paper is organized as follows: Section~\ref{ETG-sample} provides a brief
outline of the CALIFA survey and basic information on the studied ETG
sample. In Sect.~\ref{short_methodology} we briefly summarize our analysis
methodology and in Sect.~\ref{results} we describe the output from
\p3d\ and auxiliary codes that is considered in the subsequent analysis. In
Sect.~\ref{discussion} we provide a comparative discussion of the main
properties of type~i and type~ii ETGs (Sects.~\ref{type-i}\&\ref{type-ii})
laying emphasis on the morphology and EW distribution of their nebular
component.  The incidence of accretion-powered nuclear activity and the
kinematical properties of gas and stars in our sample ETGs are cursorily
discussed in Sects.~\ref{AGN} and Sect.~\ref{dis:kinematics}, respectively.
Finally, in Sect.~\ref{biases} we briefly comment on how observational bias
stemming from an EW detection threshold for faint nebular emission could
affect spectroscopic classifications of ETGs.  The main results and
conclusions from this study are summarized in Sect.~\ref{summary}.

The Appendix provides theoretical predictions from the literature and our
own evolutionary synthesis code on the time evolution of the \ewha\ for a
pAGB-dominated stellar component (Sect.~\ref{pAGB}). Additionally, in
Sect.~\ref{meth} we discuss the main components of \p3d, and the
methodology employed for the determination of emission-line fluxes and
their uncertainties, as well as the derivation of the radial distribution
of various quantities of interest.  This article is supplemented by the
main output from \p3d\ for our sample ETGs (Sect.~\ref{ETGsample}) and a
comparison sample of composite/star-forming galaxies from CALIFA
(Sect.~\ref{LTGsample}), which is meant to illustrate the variation of the
\ewha, $\tau$ and diagnostic line ratios from early-type towards late-type
galaxies.
\label{dis:kinematics}

\section{Data sample \label{ETG-sample}}

The CALIFA survey \citep[][see also S\'anchez 2014a for a
  review]{Sanchez2012} aims at a systematic study of a representative
sample of the Hubble-type galaxy population in the local Universe ($z\leq
0.03$).  The CALIFA sample \citep[600 galaxies,][]{Walcher14} has been
selected such as to ensure optimal statistics for the different Hubble
types, and to take maximum advantage of the large hexagonal FoV
(74\arcsec$\times$62\arcsec, sampled by 331 fibers 36 of which are
dedicated to sky background subtraction) of the PMAS/PPak spectrograph
\citep{Roth05,Kelz06} mounted at the 3.5m telescope at the Calar Alto
observatory.  For each sample galaxy, the CALIFA IFS data combine 2 to 3
individual dithered exposures in two different instrument setups: the
low-spectral resolution setup (V500; $R\sim 850$) and the medium-resolution
setup (V1200; $R\sim 1700$) with a spectral coverage between 3750 -- 7500
\AA\ and 3700 -- 4200 \AA, respectively. After reduction with the CALIFA
IFS data processing pipeline \citep[version 1.3c for the data set used
  here; see][for details]{Husemann2013}, the data are provided to the
CALIFA collaboration interpolated within a
78\arcsec$\times$72\arcsec\ grid, flux-calibrated at a precision better
than $\sim$15\% and corrected for foreground Galactic extinction.
Additionally, the CALIFA data processing pipeline provides from its version
v1.3c onward error statistics for each spaxel and a detailed quality
control (e.g., spectral masks with spurious features, such as local
residuals in the sky- or cosmic-ray removal).  Regular public data releases
of fully reduced CALIFA IFS data, the latest one \citep[DR2; 200 galaxies
  --][]{GarciaBenito15} in fall 2014, underscore the legacy potential of
the CALIFA survey and its general value to the astronomical community.

Following its first science publication in K12, the CALIFA collaboration
has completed several studies dedicated to, e.g., aperture effects
\citep{IP13} and the spatial resolution of IFS data \citep{Mast14}, stellar
age gradients in galaxies \citep{Perez13,Rosa14a,SB14} and the
uncertainties in their determination from spectral fitting \citep{cid14},
the mass-metallicity relation \citep{Sanchez13,Rosa14b}, the presence of
universal metallicity gradients in galaxies
\citep{Sanchez2014a,Sanchez2014b}, the spatial distribution and excitation
mechanisms of nebular emission in ETGs and other LINER galaxies
\citep[P13,][]{Singh13}, and devised a new semi-empirical metallicity
calibration based on the largest hitherto compiled sample of H{\sc ii}
regions in late-type galaxies \citep{Marino13}. Several parallel ongoing
studies within the collaboration explore various properties of present-day
Hubble-type galaxies, such as their stellar and gas kinematics
\citep[][Falc\'on-Barroso et~al., in prep.]{GL13,BB14}.

This study is based on V500 IFS datacubes for 20~E and 12~S0/SA0 galaxies
in the local volume ($<$150 Mpc) and comprises essentially all ETGs
observed by CALIFA at the onset of this project (mid 2012) with our pilot
study in K12.  The low-resolution (V500) CALIFA IFS data have the advantage
of covering the entire optical spectral range, thereby permitting flux
determinations of several emission lines and their ratios, and spectral
fitting of single-spaxel spectra down to a surface brightness level $\mu
\simeq 23.6$ $g$ \sbb\ (cf, e.g., K12), allowing for the study of the
extranuclear component of ETGs out to typically $\ga$2 $r$ band
Petrosian\_50 radii \rPetr, taken from SDSS.

In P13 we studied these 32 ETGs (Table~\ref{tab:sample}) only with respect
to the radial distribution of their EW(\ha), $\tau$ and BPT ratios, without
discussing the 2D properties of their nebular and stellar components, which
is the goal of the present article.

\section{Outline of the analysis methodology \label{short_methodology}}
The CALIFA IFS data were processed spaxel-by-spaxel with \p3d, a pipeline
developed by us with the purpose of automated spectral fitting and
post-processing of flux-calibrated IFS data cubes.  This pipeline (see
Fig.~\ref{fig:Porto3D} for a schematic overview) combines a suite of
modules written in the MIDAS\footnote{Munich Image Data Analysis System,
  provided by the European Southern Observatory (ESO).}  script language
and in GNU~Fortran~2008, with peripheral modules using CFITSIO and PGplot
routines.  The spectral fitting module of \p3d\ invokes the publicly
available (v.4) version of the population synthesis code
\SL\ \citep{cid05}.  \p3d\ enabled the first science publication of the
CALIFA collaboration in K12 and since then has undergone several upgrades.
Its current version (v.2), used in P13 and this study, incorporates several
additions, in particular, it provides estimates of uncertainties in
emission-line fluxes and EWs.

The pipeline consists of three main modules: The first one (\rem{m.1})
performs a data quality assessment and initial statistics, and extracts
from the flux-calibrated CALIFA IFS data cubes individual spectra, which in
turn are transformed into a format suitable for fitting with \SL.  Module
\rem{m.2} is intended to the extraction of emission lines, after
subtraction of the best-fitting stellar fit, and the determination of
emission-line fluxes and kinematics. It also computes various secondary
quantities from the \SL\ models (e.g., luminosity- and mass-weighted
stellar age and metallicity, the luminosity fraction of stars younger than
5 Gyr, and others).  The third module, \rem{m.3}, determines the Balmer
line luminosities implied by the \lyc\ output from the best-fitting
population vector.  The predicted \ha\ luminosity is in turn compared with
the observed one in order to assess via the $\tau$ ratio (K12,P13) the
consistency of the results with the pAGB photoionization hypothesis.

Each ETG has been processed with \p3d\ twice, based on two different
libraries of Simple Stellar Population (SSP) models that were used for
spectral modeling with \SL.  The emission-line maps produced by these two
runs were in turn combined by a decision-tree routine, which permits check
of the soundness and subsequent error-weighted averaging of values
spaxel-by-spaxel.  \p3d\ is supplemented by a routine for computation of
radial profiles of various quantities of interest (e.g., emission-line
intensities and EWs, $\tau$ and BPT ratios). This add-on module is
essentially an adaptation of the isophotal annuli surface photometry
technique \trem{iv} by \cite[][hereafter P02]{P02}.


\begin{table*}[t]
\caption{General properties of the sample ETGs.}
\tiny
\begin{tabular}{lllccrcrcccclc}
Name   \pano  & Morph. &  T  &$\alpha$(J2000)&$\delta$(J2000) &D           &  m$_{\rm r}$ & \rPetr\ & f$_{\rm H\alpha}$&L$_{\rm H\alpha}$& \mstar(e/p) & $\langle \tau \rangle$ & SC   &  KC  \\
(1)    \kato  &  (2)   &(3)  &  (4)          &    (5)         &(6)         &   (7)      & (8)      &    (9)  &   (10)            & (11)         & (12)                   & (13) & (14) \\
\hline \pano 
\object{UGC 5771} & S0/a    &i   &	10:37:19.33   &   +43:35:15.31   &   106.4      &  13.23  &  7.3   &251 & 34.0  & 11.560/11.398 & 0.69 & L               & r$\parallel$r \\
\object{NGC 4003} & SB0     &i   &	11:57:59.03   &   +23:07:29.63   &    96.6      &  13.21  &  9.6   &455 & 50.8  & 11.378/11.218 & 0.57 & C               & r$\parallel$p \\
\object{UGC 8234} & S0/a    &i   &	13:08:46.50   &   +62:16:18.10   &   116.1      &  12.82  &  5.2   &99  & 15.9  & 11.270/11.131 & 1.09 & L               & r$\parallel$n \\
\object{NGC 5966} & E       &i   &      15:35:52.10   &	  +39:46:08.05 	 &    69.0      &  12.52  &  9.4   &206 & 117.2 & 11.311/11.155 & 1.26 & L               & r$\parallel$o \\
\object{UGC 10205}& Sa      &i   &	16:06:40.18   &   +30:05:56.65   &    97.6      &  13.16  & 12.2   &561 & 63.9  & 11.375/11.221 & 0.20 & C/L             & r$\parallel$p \\ 
\object{NGC 6081} & S0      &i   &	16:12:56.85   &   +09:52:01.57   &    79.6      &  12.86  &  8.2   &249 & 18.9  & 11.493/11.331 & 1.07 & L               & r$\parallel$p \\ 
\object{NGC 6146} & E?      &i   &	16:25:10.32   &   +40:53:34.31   &   127.4      &  12.49  &  7.2   &180 & 34.9  & 11.880/11.720 & 1.86 & L               & r$\parallel$p \\
\object{UGC 10695}& E       &i   &	17:05:05.57   &   +43:02:35.35   &   120.1      &  13.19  & 10.7   &274 & 47.3  & 11.493/11.335 & 0.90 & L(S3)           & n$+$o         \\
\object{UGC 10905}& S0/a    &i   &	17:34:06.43   &   +25:20:38.29   &   114.1      &  13.03  &  7.1   &217 & 33.8  & 11.641/11.487 & 0.98 & L               & r$\parallel$p \\
\object{NGC 6762} & S0/a    &i   &      19:05:37.09   &   +63:56:02.79 	 &    44.9      &  13.19  &  5.8   &236 & 56.7  & 11.678/11.524 & 0.77 & L               & r$\parallel$r \\
\object{NGC 7025} & Sa      &i   &	21:07:47.33   &   +16:20:09.22   &    70.7      &  12.20  &  9.2   &604 & 36.1  & 11.655/11.496 & 1.67 & L               & r$\parallel$p \\ 
\hline \hline
\object{NGC 1167} & SA0     &i+  &	03:01:42.33   &   +35:12:20.21   &    66.2      &  12.19  & 13.6   &827 & 43.3  & 11.572/11.410 & 0.87 & L(S3)$\bullet$1 & r$\parallel$p \pano \\
\object{NGC 1349} & S0      &i+  &	03:31:27.51   &   +04:22:51.24   &    87.7      &  13.09  & 11.7   &467 & 42.9  & 11.341/11.183 & 0.89 & L               & r$\parallel$p \\
\object{NGC 3106} & S0      &i+  &	10:04:05.25   &   +31:11:07.65   &    90.1      &  12.69  & 11.0   &651 & 63.0  & 11.474/11.314 & 0.60 & L               & r$\parallel$p \\
\hline \hline
\object{UGC 29}   & E       &ii  &      00:04:33.74   &   +28:18:06.20   &    118.5     &  13.44  & 8.9    &33  & 5.6   & 11.362/11.203 & 4.98 & L               & n$+$n \pano   \\
\object{NGC 2918} & E       &ii  &	09:35:44.04   &   +31:42:19.67   &    98.1      &  12.56  & 8.3    &60  & 6.9   & 11.501/10.340 & 4.87 & L               & r$\parallel$n \\
\object{NGC 3300} & SAB(r)0 &ii  &	10:36:38.44   &   +14:10:15.97   &    48.0      &  12.45  & 10.9   &39  & 1.1   & 11.012/10.858 & 19.8 & L               & r$\parallel$n \\
\object{NGC 3615} & E       &ii  &	11:18:06.65   &   +23:23:50.36   &    98.3      &  12.50  & 6.9    &35  & 4.0   & 11.690/11.523 & 10.8 & L               & r$\parallel$n \\
\object{NGC 4816} & S0      &ii  &	12:56:12.14   &   +27:44:43.71   &   102.6      &  12.91  & 12.7   &35  & 4.4   & 11.530/11.367 & 18.6 & L               & p$\parallel$n \\ 
\object{NGC 6125} & E       &ii  &	16:19:11.53   &   +57:59:02.89   &    72.2      &  12.13  & 9.1    &52  & 3.2   & 11.600/11.437 & 23.4 & L               & p$+$n         \\ 
\object{NGC 6150} & E?      &ii  &	16:25:49.96   &   +40:29:19.41   &   126.0      &  13.12  & 7.6    &20  & 3.8   & 11.660/11.496 & 10.9 & L$\bullet$2     & r$\parallel$n \\
\object{NGC 6173} & E       &ii  &	16:29:44.87   &   +40:48:41.96   &   126.9      &  12.33  & 11.0   &45  & 8.6   & 12.161/11.999 & 15.6 & L               & r$\not\parallel$n \\
\object{UGC 10693}& E       &ii  &	17:04:53.01   &   +41:51:55.76   &   120.5      &  12.70  & 9.3    &34  & 5.9   & 11.731/11.572 & 11.7 & L               & p$\parallel$n \\
\object{NGC 6338} & S0      &ii  &	17:15:22.97   &   +57:24:40.28   &   117.5      &  12.56  & 11.3   &178 & 29.4  & 11.870/11.707 & 2.21 & L               & r$\not\parallel$n \\
\object{NGC 6411} & E       &ii  &	17:35:32.84   &   +60:48:48.26   &    57.6      &  12.09  & 11.3   &97  & 3.9   & 11.363/11.203 & 8.62 & C/L             & p$\parallel$p \\
\object{NGC 6427} & S0      &ii  &	17:43:38.59   &   +25:29:38.18   &    51.3      &  12.71  & 4.9    &68  & 2.1   & 11.141/10.985 & 7.60 & L               & r$\parallel$n\\
\object{NGC 6515} & E       &ii  &	17:57:25.19   &   +50:43:41.24   &    99.0      &  12.87  & 9.2    &87  & 10.2  & 11.474/11.316 & 2.46 & L/S             & n$+$n \\
\object{NGC 7194} & E       &ii  &	22:03:30.93   &   +12:38:12.41   &   110.7      &  12.87  & 6.6    &30  & 4.4   & 11.698/11.536 & 12.9 & L               & r$\parallel$n\\
\object{NGC 7236} & SA0     &ii  &	22:14:44.98   &   +13:50:47.46   &   108.2      &  13.72  & 3.4    &35  & 5.0   & 11.565/11.403 & 8.16 & L               & r$\not\parallel$n\\
\object{UGC 11958}& SA0     &ii  &      22:14:46.88   &   +13:50:27.13   &   108.3      &  13.35  & 14.2   &153 & 21.4  & 11.776/11.612 & 2.67 & L$\bullet$3     & n$+$n\\ 
\object{NGC 7436B}& E       &ii  &	22:57:57.54   &   +26:09:00.01   &   100.6      &  12.80  & 10.0   &94  & 11.4  & 11.867/11.708 & 6.87 & L               & r$\parallel$n\\
\object{NGC 7550} & SA0     &ii  &	23:15:16.07   &   +18:57:41.22   &    69.2      &  12.29  & 11.6   &267 & 15.3  & 11.555/11.394 & 3.02 & L               & n$+$r \\
\hline \hline
\object{IC 540}   & S       & -- &      09:30:10.33   &   +07:54:09.90   &    31.9      &  13.61  &  8.9   &401 & 4.9   & 10.152/10.004 & 0.36 & S               & r$\parallel$p \pano \\
\object{UGC 8778} & S?      & -- &      13:52:06.66   &   +38:04:01.27   &    52.6      &  13.56  &  9.6   &590 & 19.6  & 10.590/10.439 & 0.31 & C/L             & r$\parallel$r \\
\object{IC 944}   & Sa      & -- &      13:51:30.86   &   +14:05:31.95   &   104.8      &  12.89  & 10.6   &530 & 70    & 11.556/10.439 & 0.58 & C/L             & r$\parallel$p \\
\object{IC 1683}  & S?      & -- &      01:22:38.92   &   +34:26:13.65   &    65.4      &  13.47  &  9.2   &1703& 87    & 10.924/10.772 & 0.18 & H               & r$\parallel$p \\
\object{NGC 4470} & Sa?     & -- &      12:29:37.77   &   +07:49:27.12   &    38.4      &  12.57  & 13.3   &8372& 147   & 10.231/10.108 & 0.03 & H               & r$\parallel$r \\
\end{tabular}\vspace{0.3cm}
{\small

$\star$\trem{ Col. 1:}  Galaxy name.\\
$\star$\trem{ Col. 2:}  Morphological classification adopted from NASA/IPAC Extragalactic Database (NED).\\
$\star$\trem{ Col. 3:}  Classification according to our study in P13, based on the mean $\tau$ ratio (\trem{Col. 12}) for the extranuclear component (type i, i+ and ii).\\
$\star$\trem{ Col. 4:}  Right ascension.\\
$\star$\trem{ Col. 5:}  Declination.\\
$\star$\trem{ Col. 6:}  Distance in Mpc adopted from NED.\\
$\star$\trem{ Col. 7:}  Apparent $r$ total magnitude from SDSS DR7.\\
$\star$\trem{ Col. 8:}  Petrosian\_50 radius in arcsec in the SDSS $r$ band.\\
$\star$\trem{ Col. 9:}  Integral \ha\ flux in units of $10^{-16}$ \uflux, derived from CALIFA IFS data.\\
$\star$\trem{ Col. 10:} Integral \ha\ luminosity in units of 10$^{39}$ \ergsec.\\
$\star$\trem{ Col. 11:} Logarithm of the ever formed and present stellar mass in \msun, as obtained from spectral modeling with \SL.\\
$\star$\trem{ Col. 12:} Mean $\tau$ ratio (from P13 in the case of ETGs).\\  
$\star$\trem{ Col. 13:} Spectroscopic classification based on the emission-line ratios (panels i\&j of Figs.~\ref{fig:bpt_n1167}, \ref{fig:bpt_u05771}--\ref{fig:bpt_n7550}
and \ref{fig:bpt_ic0540}--\ref{fig:bpt_n4470}) of the nuclear (\rr$\leq$3\farcs8) CALIFA spectrum. 
The notation is as follows: {\rem H}: H{\sc ii}/star-forming, {\rem C:} composite (cf description of panels i\&j of Fig.~\ref{fig:bpt_n1167}), 
{\rem L}: LINER, {\rem S}: Seyfert. 
Whenever available, the spectroscopic classification from NED is included within brackets, and the index following the $\bullet$ symbol points to additional 
information available at NED on the morphology of the radio continuum: (1) Radio jet, (2) Narrow-angle tail radio source, (3) FR\,I source.\\
$\star$\trem{Col. 14:} Tentative 3-character kinematical classification: {\bf r}: rotation-dominated kinematics, {\bf p}: perturbed rotational pattern, 
{\bf o}: gas outflow, and {\bf n}: no distinguishable pattern or pressure-supported stellar kinematics. 
The first and third character refer, respectively, to the kinematics of stars and gas, and the second one indicates in the 
case of stellar rotation (r or p) whether the kinematic and photometric major axes are roughly aligned ($\parallel$) or not ($\not\parallel$).
The + symbol marks ETGs where stellar rotation is not detected (n) or having a nearly circular-symmetric morphology, thus uncertain
photometric major axis. \\

Note that the galaxies \object{NGC 7025}, \object{NGC 7236} and
\object{NGC 7436B} are overlapping over a substantial fraction of
their area with extended foreground/background sources that, even
after subtraction, are likely introducing uncertainties of $\sim$10\%
in the quoted \ha\ fluxes.}
\label{tab:sample}
\end{table*}

\section{Results \label{results}}
Figure~\ref{fig:bpt_n1167} shows a representative example of the
information extracted and used in the discussion of our ETG sample. This
figure along with the results obtained for the remaining sample galaxies
(Figs.~\ref{fig:bpt_u05771}--\ref{fig:bpt_n7550}; Appendix~\ref{ETGsample})
illustrate the considerable diversity of local ETGs with respect to
morphology, kinematics and physical properties of their warm interstellar
medium. The~2D maps (panels \rem{a}--{\rem{f}~\&~\rem{n}) cover an area of
  {\small 78\arcsec$\times$72\arcsec} and are displayed in astronomical
  orientation (north is up and east to the left).

Panel \rem{a}: Logarithmic representation of the emission-line free stellar
continuum flux (in units of $10^{-16}$ \uflux) in the spectral range
between 6390 $\AA$ and 6490 $\AA$. The cross (in panels \rem{a}--\rem{f}
and \rem{n}) marks the maximum intensity of the stellar emission, the
morphology of which is further delineated by the overlaid contours at 32,
10, 3, 2.3 and 1.3\% of the peak intensity. The linear scale in kpc
labeling the 20\arcsec horizontal bar to the upper-right has been computed
from the adopted distance to the source (Col.~6 in Table~\ref{tab:sample}).
Regions being contaminated by probable foreground or background sources,
therefore excluded from the analysis, are left blank.

Panel \rem{b}: \ha\ flux map in units of $10^{-16}$ \uflux.

Panel \rem{c}: \ewha\ map in \AA.

Panel \rem{d}: Stellar velocity field after correction for systemic
velocity, as derived from \SL\ fits to individual spaxels. The right-hand
side vertical bar indicates the dynamical range of the image in \kmsec, and
the red circle centered on the linear scale bar illustrates the effective
angular resolution (FWHM$\approx$3\farcs8 for data cubes processed with
version 1.3c of the CALIFA data reduction pipeline) of our IFS data. The
blue lines show the photometric minor axis.

Panel \rem{e}: Ionized gas velocity field, as obtained by averaging the
\ha\ and [N{\sc ii}]$\lambda$6584 velocity maps.

Panel \rem{f}: Subdivision of the \ewha\ map into three intervals, meant to
help the eye to distinguish between regions where the observed \ewha\ is
consistent with pure pAGB photoionization (0.5--2.4~\AA; light blue), an
additional gas excitation source is needed to account for the observed
\ewha\ ($>$2.4~\AA; orange), and where the \ewha\ is lower than that
predicted from pAGB photoionization models ($\leq$0.5~\AA; dark blue; cf
Sect.~\ref{pAGB}), thus \lyc\ photon escape is likely important. The
fraction in \% of the spectroscopically studied area where the observed
EW(\ha) implies \lyc\ leakage, is consistent with pure pAGB
photoionization, or an extra ionization source is required is indicated in
the upper-right of this panel.  Note that the adopted lower bound of
0.5~\AA\ ($\equiv$\pAGBmin) corresponds to the \ewha\ expected for an
instantaneous burst with an age of $\sim$1~Gyr (cf evolutionary synthesis
models in Sect.~\ref{pAGB}), i.e. it is rather representative of ETGs
having undergone significant recent stellar mass growth.  Since the
\ewha\ expected for a typical ETG (age $\geq$5~Gyr) is $\geq$1~\AA, the
condition \ewha$\geq$0.5~\AA\ is rather a generous minimum requirement for
the validity of the pAGB photoionization hypothesis, provided that
\lyc\ escape is negligible.  Panels \rem{g}\&\rem{h}: \ha\ intensity,
normalized to its peak value, and \ewha\ as a function of the photometric
radius \rr\ (\arcsec). Determinations based on single spaxels (\sisp)
within the nuclear region, defined as twice the angular resolution of the
IFS data (i.e. for \rr$\leq$3\farcs8) and in the extranuclear component are
shown with red and blue circles, respectively. Open squares (green color)
correspond to the average of individual \sisp\ determinations within
isophotal annuli (\isan; cf Sect.~\ref{radial_profiles}) with vertical bars
illustrating the $\pm$1$\sigma$ scatter of data points.  The light-shaded
area illustrates the \ewha\ range that can be accounted for by pAGB
photoionization for ages between $\sim$0.1 Gyr and $\sim$13~Gyr for the
full range in metallicity covered by SSP models from \citet[][hereafter
  BC03]{bru03}.  This range includes a brief minimum (0.1--0.8~\AA) at the
onset of the pAGB phase and an upper range of 3.4~\AA\ for a super-solar
metallicity of 2.5\,\zsun\ (0.1--3.4 \AA; cf Fig.~\ref{fig:rebetiko} in
this paper and Fig.~2 in \citet{cid11}).  However, for reasonable
assumptions on the age (1--12 Gyr) and metallicity ($\sim$\zsun) of stellar
populations in ETGs (Sect.~\ref{pAGB} for details), the range of expected
\ewha's can be narrowed down to within \pAGBmin=0.5~\AA\ and
\pAGBmax=2.4~\AA\ (dark-gray strip) with a typical value around
1~\AA\ ($\equiv$\pAGBmean) in the age interval between $\sim$6~Gyr and
$\sim$11~Gyr.

Panels \rem{i}\&\rem{j}: \sisp\ determinations for the BPT-VO diagrams on
the \tln2ha\ vs \tlo3hb\ and \tlsiiha\ vs \tlo3hb\ diagnostic emission-line
ratios planes.  The meaning of symbols is identical to those in panels
\rem{g--h}.  The loci on the BPT diagrams that are characteristic of AGN
and LINERs, and that corresponding to photoionization by young massive
stars in H{\sc ii} regions are indicated, with demarcation lines from
\citet[][dotted curve]{Kauffmann03}, \citet[][solid curve]{Kewley01} and
\citet[][dashed line]{Schawinski2007}. For the sake of comparison, the grid
of thin-gray lines roughly at the middle of each diagram depicts the
parameter space that can be accounted for by pure shock excitation, as
predicted by \cite{Allen2008} for a magnetic field of 1~$\mu$G, and a range
of shock velocities between 100 and 1000 \kmsec, for gas densities between
0.1 and 100 cm$^{-3}$.

Panels \rem{k}\&\rem{l}: \tlo3hb\ and \tln2ha\ ratio as a function of \rr,
as inferred from \sisp\ and \isan\ determinations. The shaded horizontal
area depicts the mean ratios for our sample (cf P13) of 0.37$\pm$0.13 for
\lo3hb\ and 0.34$\pm$0.26 for \ln2ha, with a standard deviation about the
mean $\sigma_{\rm N}$ of 0.02 and 0.05, respectively.

Panel~\rem{m}: $\tau$ ratio radial distribution, without and after
correction for intrinsic stellar extinction (vertical lines connecting
symbols at equal \rr). The locus in this diagram where the \lyc\ output
from the pAGB component precisely accounts for the observed \ha\ luminosity
is depicted by the thick horizontal line at log($\tau$)=0, and the dashed
line at log($\tau$)=0.3 marks the boundary between pAGB photoionization and
\lyc\ escape (cf P13).

Panel \rem{n}: Luminosity contribution (\lfrac[\%]) of stellar populations
younger than 5~Gyr at 5150~\AA\ (normalization $\lambda$).

In the Appendix~\ref{LTGsample} we additionally include five
intermediate-to-late type CALIFA galaxies that were processed with \p3d\ in
the same manner as ETGs, except for using only BC03 SSPs for spectral
fitting. All but one of these systems are classifiable by the BPT ratios in
their central part and/or extranuclear component as {\sl composite} or
SF-{\sl dominated} (C and H, respectively, in the notation of Table~1;
col. 13). These non-ETG galaxies (not further discussed here) are only
meant to illustrate how several quantities of interest, and their
uncertainties, vary from almost \wim-devoid type~ii ETGs towards
SF-dominated late-type galaxies with an \ewha\ several times larger than
\pAGBmax.  For example, it is interesting to note how the fractional area
in which extra (non-pAGB) photoionization needs be postulated (yellow
regions in panel \rem{f}) strikingly increases from ETGs towards ``later''
galaxy types, reaching in the case of \object{NGC 4470} $\sim$90\% of the
total spectroscopically studied area.  This trend is accompanied by a
predominance of stars younger than 5~Gyr to the optical luminosity (panel
\rem{n}).  Noteworthy is also the reduction of the mean radial \lo3hb\ and
\ln2ha\ ratios -- both now offset by $\ga$--0.5~dex relative to the
LINER-typical ratios of ETGs -- placing \object{NGC 4470} throughout its
extent into the `composite/SF' regime of the BPT plane.

\plotETG[arqO=NGC_1167,arqT=n1167,BPTs=bpt_n1167,name=NGC 1167,type=i+,morp=S0,tauR=0.9,dist=66.2,Magr=-21.91,lumH=43.3,Mass=11.410,desX=0,desY=0] \\

\section{Discussion \label{discussion}}

\subsection{2D properties of gas and stars in the sample ETGs \label{dis0}}
The \lfrac\ maps (Figs.~\ref{fig:bpt_n1167}n--\ref{fig:bpt_n7550}n) echo
the well established fact that ETGs are dominated by an evolved stellar
component throughout their optical extent. Even though there is a trend for
an increasing luminosity contribution from stars younger than 5~Gyr towards
the galaxy periphery, in agreement with the conclusions by, e.g.,
\citet{Rosa14a}, panels \rem{n} reveal, just like the inwardly increasing
mass-weighted stellar age \tmass\ (cf Fig.~\ref{fig:tmass}), that the bulk
of the stellar component in ETGs has formed several Gyr ago.

Another insight from the \ha\ and \ewha\ maps (panels \rem{b}\&\rem{c}) is
the presence of a diffuse, low-surface brightness \wim\ component over
virtually the entire optical extent of our sample ETGs. Its co-spatiality
with the evolved underlying background is consistent with the notion that
the diffuse ionizing field from the latter is an important driver of
extranuclear \wim\ emission.
Following our pilot study in K12, this issue was addressed in more detail
and on the basis of the robust statistics from this sample in P13, where we
concluded that, in about 40\% of ETGs (type~i), the properties of the
extranuclear \wim\ are indeed compatible with the pAGB photoionization
hypothesis.  The main lines of arguing in those studies were based on a
combined analysis of the \ewha\ and the $\tau$ ratio.

Regarding the first quantity, one can read off Fig.~\ref{fig:rebetiko} that
the photoionization by pAGB stars can reproduce the narrow range of EWs
between \pAGBmin\ (0.5 \AA) and \pAGBmax\ ($\sim$2.4 \AA). This range can
be further limited to $\simeq$1~\AA\ ($\equiv$\pAGBmean) when taking into
account the typical \tmass\ (7 -- 11 Gyr) of ETGs, and making the
reasonable assumption of a nearly-solar metallicity (see discussion in
Sect.~\ref{pAGB}).  Obviously, the condition
\pAGBmin$\leq$\ewha$\leq$\pAGBmax\ ensures compliance of observations with
predictions for pure pAGB models.

In passing we note that we do not invoke the presence or absence of LINER
characteristics as an argument for or against pAGB photoionization, since
it is known that they can be due to a variety of mechanisms, including
shocks and starburst-driven outflows \citep[see, e.g.,][for a recent
  observational review]{Sharp2010}.  For example, the biconical
\wim\ component protruding several kpc away from the nucleus of \object{NGC
  5966} in NE-SW direction (Fig.~\ref{fig:bpt_n5966}; see K12 for a
detailed discussion) shows throughout LINER-typical emission-line ratios
and an \ewha\ that is compatible with pure pAGB photoionization.
Notwithstanding this fact, the morphology of this kinematically decoupled
\wim\ component is suggestive of large-scale shock excitation associated to
gas outflow.  The absence of ongoing SF throughout this ETG lends support
to the idea that its gas outflow is powered by an AGN hosted in its LINER
nucleus.

As for the $\tau$ ratio, a radial constancy around unity would ensure that
pAGB stars are capable of supplying the ionizing photon budget that is
necessary for sustaining the observed \ha\ emission without need for any
additional excitation source. The latter case would correspond to $\tau<1$,
whereas a $\tau>1$ implies the escape of a fraction \plf=1-$\tau^{-1}$ of
the Lyman continuum radiation produced by the pAGB component
(Sect.~\ref{intro}).  As argued in \cite{cid11} a relation between the
\ewha\ and $\tau^{-1}$ is to be expected, and it was found indeed from an
analysis of Sloan Digital Sky Survey (SDSS) data (see their Fig.~5).
Likewise, a tight anticorrelation between $\tau$ and \ewha\ over nearly 3
dex, with the pAGB ``equilibrium'' line of $\tau\simeq 1$ corresponding to
\ewha$\simeq$1~\AA\ ($\equiv$\pAGBmean) was documented from the analysis of
CALIFA IFS data in P13 (see their Fig.~2a).

One may superficially argue that EW(\ha) and $\tau$ reflect essentially one
and the same quantity, they therefore can be used interchangeably in
assessing the validity of pAGB photoionization hypothesis.  However, an
appreciation of the a) geometry-dependent line-of-sight dilution effect of
nuclear EWs in triaxial stellar systems and b) the large \lyc\ escape
fraction in the majority of ETGs (see discussion in P13) call this
simplistic view into question. For example, a consequence of the geometric
dilution is that a radially constant \ewha\ in the range between
\pAGBmin\ and \pAGBmax\ can only be regarded as a proof for the validity of
the pAGB photoionization hypothesis in the specific case of an oblate
stellar system seen nearly face-on, or when in a triaxial stellar system
the line-emitting \wim\ is uniformly mixed with the EW-diluting stellar
background. Whenever the \wim\ occupies a smaller volume than the stars,
the intrinsic \ewha\ within a volume element is greater than the observed
(projected) value.  This cautionary note is essential for a proper
evaluation of panels \rem{f} where regions satisfying the condition
\pAGBmin$\leq$\ewha$\leq$\pAGBmax\ are depicted in open-blue color: pAGB
photoionization can be regarded as the dominant gas excitation source in
these regions only as long as the \wim\ is co-spatial with the stars and
\lyc\ escape is throughout negligible.  Would any of these (far from
self-evident) assumptions be invalid, then an additional excitation source
(e.g., AGN or diffuse SF) would need to be postulated.  In fact, comparison
of panels \rem{b} \& \rem{c} shows the importance of the line-of-sight EW
dilution effect in several of our sample galaxies. \object{NGC 5966} offers
again an illustrative example: The biconical \ha\ emission in this system
has its maximum at the peak of the stellar surface density (panel \rem{b}),
whereas the opposite is the case for the \ewha\ (panel \rem{c}).  Note that
a spatial anti-correlation between emission-line fluxes and EWs has been
documented and discussed also in other triaxial systems, such as blue
compact dwarf galaxies \citep[see, e.g.,][]{P02}.

In P13 we showed that, judging from their \ewha\ and $\tau$ profiles in
their extranuclear zones (\rr$\geq$4 \arcsec; blue dots in panels
\rem{g}--\rem{m}), ETGs span a broad, continuous sequence that can be
tentatively grouped in two main classes, based on radial \ewha\ profiles
(see schematic representation in Fig.~\ref{fig:ETG-types}).  Type~i ETGs
(14 galaxies) are characterized by a nearly constant \ewha\ of
$\ga$1~\AA\ out to their periphery, with the exception of a few systems
where the EW shows a steep outer increase to \ewha$\gg$\pAGBmax. Such
systems, to which \citet{G14b} assign the notation i+, are subject of a
recent parallel CALIFA study \citep{G14a}, which reveals that the EW excess
is due to low-level star-forming activity in faint embedded spiral-like
features.  Type~ii ETGs (18 galaxies), on the other hand, display very low
($\leq$0.5 \AA) EWs in their cores and typically positive EW gradients,
approaching an \ewha$\simeq$\pAGBmean\ only in their outermost periphery
(\rr$\ga$2 \rPetr).

\begin{figure}
\begin{picture}(8.6,4.2)
\put(0.0,0.0){\includegraphics[width=9.00cm, clip=true, viewport=30 110 750 440]{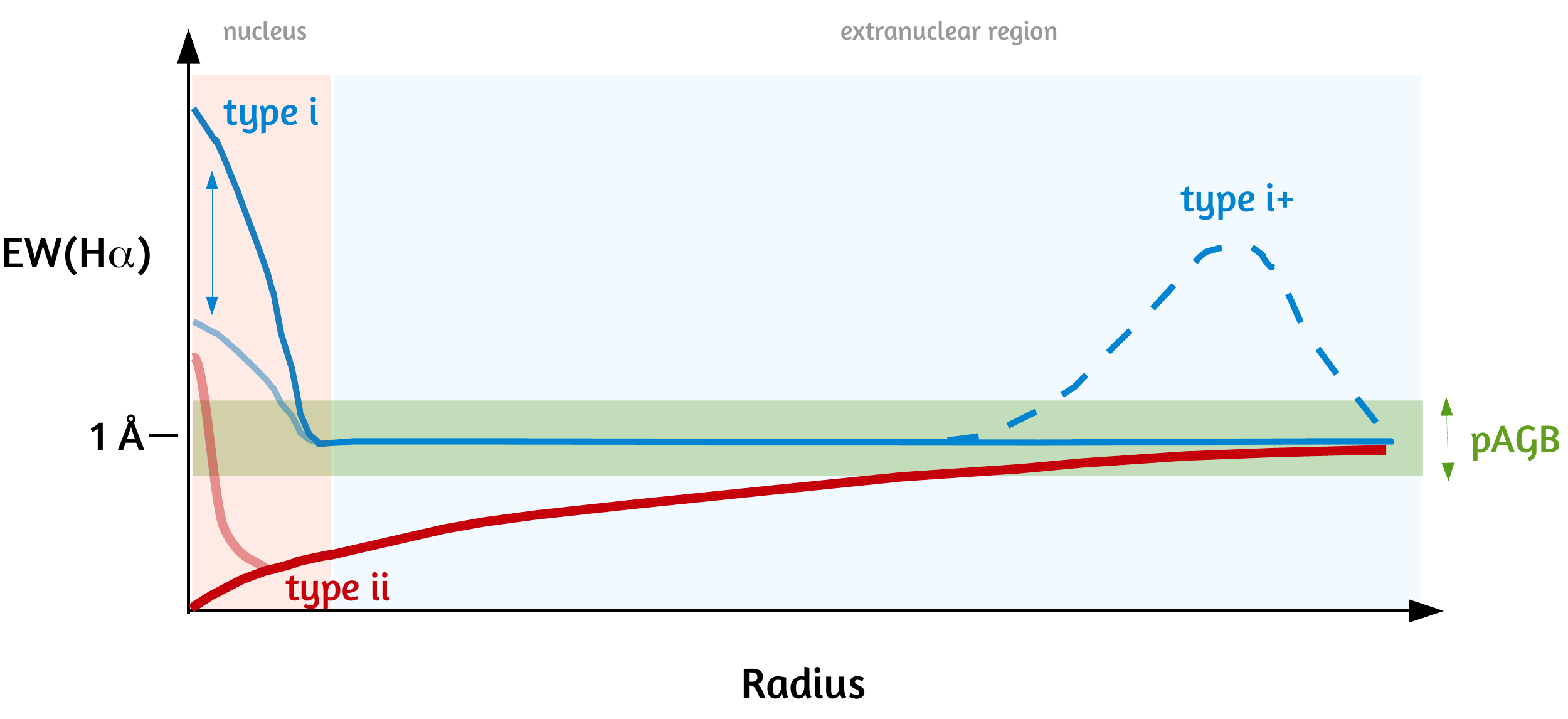}}
\end{picture}

\caption[]{\color{black} Schematic representation of the two main classes
  of ETGs, as defined in P13.  The \ewha\ profiles of \trem{type~i} ETGs
  (blue color), such as, e.g., \object{NGC 1167}, show in their
  extranuclear component (beyond the radius range depicted by the vertical
  shaded area; typically $\sim$2 kpc in our sample) nearly constant values
  within the narrow range between \pAGBmin\ and \pAGBmax\ (green-color
  horizontal strip), with a mean value of typically $\sim$1~\AA.  A few of
  these systems (labeled i+, blue-dashed curve; e.g., \object{NGC 1349})
  additionally show a steep \ewha\ increase above \pAGBmax\ in their
  periphery ($>$1\,effective radius), which, as we discuss in
  \citet{G14b,G14a} (see also Sect.~\ref{type-i}), is owing to low-level
  star-forming activity.  By contrast, \trem{type~ii} ETGs (red curve) show
  a centrally very low ($\leq$0.5~\AA) mean \ewha, increasing then smoothly
  to $\ga$\pAGBmin\ at their periphery (e.g., \object{NGC 6411}). Regarding
  their nuclear properties, both ETG types display a considerable
  diversity, from systems with virtually \wim-evacuated cores ($<$0.5~\AA)
  to galaxies hosting a compact central core with an \ewha\ of a few
  \pAGBmax.  The profiles in light-blue and light-red color are added to
  illustrate the spread in the nuclear \ewha's of our sample ETGs. }
\label{fig:ETG-types}
\end{figure}

\subsection{Nebular emission in type~i ETGs \label{type-i}}

With regard to most type~i ETGs in our sample, our results are consistent with
(yet, recalling our cautionary notes in the previous section, no proof for)
the hypothesis that the diffuse ionizing photon field from the pAGB component
is the main driver of extranuclear \wim\ emission.  Panels \rem{h} in
Fig.~\ref{fig:bpt_n1167} and Figs.~\ref{fig:bpt_u05771}-\ref{fig:bpt_n7550}
show that most \isan\ determinations for type~i ETGs satisfy the condition
\pAGBmin$\leq$\ewha$\leq$\pAGBmax\ (shaded area in panels \rem{f}), with most
data points populating the EW strip around \pAGBmean. Consistently with this
trend, most type~i ETGs show a $\tau \simeq 1$, with all data points located
beneath $\tau=2$ (dashed horizontal line in panels \rem{m}).

\color{black} Regarding the subclass of type~i+ ETGs, two of them are
morphologically classified as S0 (\object{NGC 1349} and \object{NGC 3106}),
the third one as SA0 (\object{NGC 1167}). As shown in \citet{G14a}, the
outer EW-enhanced zone of these systems (panel \rem{c}) is due to
photoionization by young stars over a fraction between $\sim$7\% and
$\sim$47\% of the galaxy area.  Note that on the radial BPT ratio profiles
(i.e. the \isan-based \o3hb\ and \n2ha\ determinations in panels
\rem{k}\&\rem{l}) the outer star-forming zone of type~i+ ETGs is clearly
separable from the LINER-typical average for ETGs (gray strip) only when
producing significant nebular emission ($\geq$6~\AA; \object{NGC 1349} and
\object{NGC 3106}; Figs.~\ref{fig:bpt_ngc1349} and
Figs.~\ref{fig:bpt_ngc3106}, respectively), yet marginally traceable in the
case of \object{NGC 1167} (Fig.~\ref{fig:bpt_n1167}) where the
\ewha\ ($\sim$3--4.5~\AA) is just above \pAGBmax.  As discussed in
\citet{G14a}, the spectroscopic BPT classification of type~i+ ETGs is
strongly dependent on the spectroscopic aperture used (or, equivalently, on
the redshift of such a source).  The presence of SF activity in the
periphery of the three type~i+ ETGs in our sample was noticed in previous
studies of the CALIFA collaboration \citep{Sanchez2014a}, even though no
explicit discussion was given to this subject due to their specific scope.
For example, the analysis underlying the aforementioned studies has
established that the outer \ha\ rim in \object{NGC 1349} is composed of
several genuine H{\sc ii} regions, that is clumpy \ha\ entities containing
a young ionizing stellar component with a luminosity fraction $\geq$20\%
and showing an \ewha$\geq$6~\AA\ \citep[see, e.g.,][]{Sanchez2014b}.

Besides the three ETGs classified type~i+, our sample contains a few
further systems witnessing the potentially relevant role of SF over an
extended radial zone or even galactic scale.  In fact, the interpretation
of pAGB photoionization dominating at all radii is challenged by singular
or contiguous EW enhancements above \pAGBmean\ in the extranuclear part of
several ETGs. One such example is the S0 galaxy \object{NGC 6081}
(Fig.~\ref{fig:bpt_n6081}c). This type~i ETG shows LINER characteristics in
its nucleus and intermediate zones (\rr$\la$10\arcsec) notwithstanding a
contiguous weak (\ewha$\la$2.5 \AA) SF rim all over its southwestern half.
Still, \isan\ determinations for the \ewha\ and $\tau$ are consistent with
a dominant contribution from pAGB photoionization, in agreement with
Fig.~\ref{fig:bpt_n6081}f which indicates that sources other than pAGB
stars dominate in less 7\% of the area of the galaxy.  Weak embedded SF
patterns are also visible in three intermediate-morphology galaxies,
\object{NGC 7025} (Fig.~\ref{fig:bpt_n7025}b\&c), \object{UGC 10205}
(Fig.~\ref{fig:bpt_u10205}b\&c) and \object{NGC 4003}
(Fig.~\ref{fig:bpt_n4003}b\&c).  In particular, \ha\ and \ewha\ maps for
\object{NGC 7025} reveal a complex, low-level SF network with an
\ewha$>$\pAGBmax, surrounding an extended (\rr$\sim$10\arcsec) LINER/AGN
core. LINER characteristics are also apparent in the intermediate zone of
\object{NGC 4003}, whereas the nuclear region and the periphery of this ETG
are dominated by SF.  This case also illustrates how the area considered in
the spectroscopic analysis can influence the spectroscopic classification,
hence the relative role ascribed to different gas excitation mechanisms:
\sisp\ (i.e. higher-S/N) data capture only the intermediate LINER zone
surrounding the SF nucleus, whereas the SF-dominated low-surface brightness
periphery of the galaxy becomes apparent on \lo3hb\ and \ln2ha\ profiles
only through \isan\ determinations including lower-S/N spaxels.

As for the nuclear properties of type~i ETGs, all except for \object{NGC
  4003} and \object{UGC 10205} fall, according to our classification in
Table~1 into the LINER locus of BPT diagrams, just like our ETG sample as a
whole.  Note that \sisp\ determinations on the \ln2ha\ vs \lo3hb\ and
\lsiiha\ vs \lo3hb\ diagrams (panels \rem{i}\&\rem{j}) place the nuclei and
intermediate zones close to the junction between the curves which demarcate
the regions of `maximum SF' and `AGN excitation'.

Our study also shows that type~i ETGs span a wide range in the \ewha\ and
spatial extent of their nuclear emission.  A few of them show prominent
nuclear \wim\ components over $\sim$6\arcsec\--10\arcsec\ (e.g.,
\object{NGC 1167}, \object{NGC 5966}, \object{UGC 5771}, \object{NGC 6146})
with a peak EW of up to $\sim$8~\AA, whereas in others the nuclear nebular
emission is rather compact ($\sim$FWHM) and faint ($\sim$\pAGBmean).  One
example of extended nuclear emission over scales of $\sim$4 kpc is seen in
\object{NGC 1167}. Besides the biconical \wim\ protruding out to $\sim$10
kpc from the nucleus of \object{NGC 5966} in a direction perpendicular to
the major axis, bipolar or unipolar \wim\ lobes on kpc-scales from the
nucleus are detected in \object{UGC 10695} and \object{NGC 6146}.  A robust
spectroscopic and kinematical analysis of these extended, faint
(\ewha$\la$\pAGBmean) features is barely possible with the available CALIFA
V500 data due to their low spectral and spatial resolution (FWHM$\geq$1 kpc
at their distance of the four ETGs), and medium S/N.  However, the absence
of appreciable star formation in these ETG nuclei renders the
interpretation of the detected \wim lobes as AGN-driven outflows plausible,
making them targets of considerable interest for follow-up studies
exploring direct evidence for accretion-powered nuclear activity in LINER
nuclei.  Spatially resolved IFS studies of high-excitation
(i.e. AGN-specific) emission lines in the nuclear and circumnuclear region
of these candidate AGN \citep[following the approach by][]{MS11}, and
supplemented by kinematical modeling of biconical gas outflows possibly
emanating from their narrow-line regions \citep[e.g.,][]{Fischer13} could
provide key insights in this respect.

\subsection{The deficit of nebular emission in type~ii ETGs \label{type-ii}}
The nature of type~ii ETGs -- systems commonly referred to as `ETGs without
nebular emission' -- is enigmatic.  These galaxies make about 60\% (18 of
32) of our sample and are classified on the condition $\langle \tau \rangle
\geq 2$ in their extranuclear component (cf P13).  Nebular emission in
these systems is only detectable after accurate fitting and subtraction of
the local underlying stellar SED, which clearly is a demanding task, given
its extreme faintness. This is particularly true for the virtually
\wim-evacuated cores of these ETGs, where the \ewha\ has its minimum,
rising then gradually to levels $\ga$\pAGBmin\ up to \pAGBmean\ in the
outermost galaxy periphery ($\sim$2\rPetr).  These positive
\ewha\ gradients, as distinctive feature of type~ii ETGs, were interpreted
by P13 as manifestation of the radius-dependent structure and physical
conditions of the gas in these systems (volume filling factor, porosity and
3D distribution; electron density and ionization parameter).  Panels
\rem{f} show that the low \ewha\ within typically more than 80\% of the
spectroscopically studied area of type~ii ETGs is incompatible with pAGB
photoionization, if case~B recombination for standard gas conditions
($n_{\rm e}=100$ cm$^{-3}$, T$_{\rm e}$=10$^4$ K) is assumed, a fact
underscoring our previous conclusion in P13 that most of the ionizing
photons generated by the pAGB component escape without being reprocessed
into nebular emission.  As a comparison, on average 65\% of the area of a
type~i ETG is consistent with the hypothesis that their gas is being
photoionized by pAGB stars.

Diagnostic emission-line ratios and their radial distribution in type~ii
ETGs have to be considered in the light of their large uncertainties, even
after averaging of \sisp\ determinations within morphology-adapted
irregular annuli (\isan). Albeit a scatter of typically 0.3~dex for
\isan\ data, one can consistently infer from panels \rem{i}\&\rem{j} that
type~ii ETGs are in practice indistinguishable from type~i ETGs by their
nearly radially constant LINER-typical BPT ratios.

As for their nuclear properties, type~ii ETGs also feature a considerable
heterogeneity: From panels \rem{b}\&\rem{c} it can be seen that in some of
these systems a compact yet extended ($\ga$3$\cdot$PSF) nuclear \ha\ excess
is robustly detected, as, for example, in \object{NGC 2918}, \object{NGC
  6150}, \object{NGC 6173}, \object{NGC 6515}, \object{NGC 7550} and, more
impressively, in \object{UGC 11958} and \object{NGC 6338}. In the majority
of type~ii ETGs, however, nebular traces of nuclear activity are almost
absent, with an \ewha\ marginally above that of the diffuse \wim\ and at
the edge of the detectability (0.1--0.3 \AA).

With regard to their peripheral zones, whereas individual \sisp\ data
points are generally subject to prohibitively large relative uncertainties
(above 100\% at EWs$<$0.3~\AA), the averaging of hundreds of such
determinations within \isan\ shows the presence of a faint substrate of
\wim\ with an \ewha$\ga$\pAGBmin.  This is augmented by determinations of
emission-line fluxes based on azimuthally binned \isan\ segments
(Fig.~\ref{fig:segmentation}) or the collapsed spectrum within each \isan,
after correction of local stellar motions (Fig.~\ref{fig:NetSpecInZones}).
It should be noted, on the other hand, that the angular resolution of our
data and the faintness of the \wim\ does not allow us to determine whether
part of the nebular emission originates from higher-density H{\sc
  ii}-region clumps within a more tenuous \wim\ component.

\subsubsection{Accretion-powered nuclear activity in ETG/LINERs \label{AGN}}
Whether or not LINER nuclei in ETGs harbor an AGN is subject of an intense
longstanding controversy, with antipodal conclusions reached, including the
radical rejection of the hypothesis of accretion-powered nuclear activity
in favor of pure pAGB photoionization (cf Introduction).  However, a first
reservation towards the latter statement comes from the established
relation between the velocity dispersion of galaxy spheroids and the mass
of their super-massive black hole (SMBH) \citep[see][for a review]{MF2001},
which is estimated to range between a few $10^7$ \msun\ and up to
$\sim$$10^{10}$ \msun.  If so, a minimum amount of gas would in principal
suffice for sustaining accretion-powered nuclear activity, perhaps in form
of a `weak', radiatively inefficient LLAGN.  As this study shows, diffuse
\wim\ is present, even in type~ii ETGs, leaving space for the AGN
hypothesis.  Indirect observational support to the latter comes from the
bi-conical/unipolar \wim\ lobes in three type~i ETGs of our sample
(Sect.~\ref{type-i}), which are consistent with AGN-driven gas outflows.
Additionally, in one type~i ETG (\object{NGC 1167}) the presence of an AGN
is indicated by a radio jet \citep[e.g.,][]{Sanghera1995,Giovannini2001}.
Also, among type~ii ETGs, radio-continuum data for \object{NGC 6150} and
\object{UGC 11958} (cf Table 1) witness the energy release by an AGN. For
example, the line-weak (\ewha$\sim$3.5 \AA) LINER nucleus of \object{UGC
  11958} powers a $\ga$100 kpc double-lobe radio continuum halo detected at
1.4 GHz with the VLA \citep{CO1991} with a large ($\sim$40 kpc) central
cavity of hot ($\sim$1 keV), pressure-driven X-ray emitting plasma
\citep{Worrall2007}, clearly indicating the energetic action by an AGN.

How to reconcile the weakness of optical emission lines in ETG/LINERs with
signatures of strong accretion-powered nuclear activity in radio- and X-ray
wavelengths poses an enigma.  As first pointed out in P13, the seemingly
trivial fact that more than one half of ETGs lacks nebular emission as a
result of an insufficient mean gas density and/or high gas porosity has
far-reaching consequences for our understanding of accretion-powered
nuclear activity in these systems: In the presence of extensive
\lyc\ photon escape, implied by the lack of gas, emission-line luminosities
and EWs, upon which estimates of accretion rates and SMBH masses rely, are
reduced by more than an order of magnitude.  An additional effect acting
towards diminishing nuclear EWs (or any photometric signs of nuclear
activity) in these triaxial stellar systems is dilution by the stellar
background along the line of sight (P13).  According to our study in P13,
the line weakness of ETG/LINER nuclei is therefore no foolproof evidence
for the weakness or absence of AGN in these systems. In the light of this
conclusion and from the evidence presented in
Sects. \ref{type-i}\&\ref{type-ii}, it is therefore conceivable that
different gas excitation mechanisms (photoionization by AGN, OB stars and
pAGB sources; shocks) -- each of them with a different relative
contribution in different radial zones -- are simultaneously at work.  An
exploration of this issue with deep, higher-spectral resolution IFS data is
apparently of great interest.

\subsection{Stellar and gas kinematics in ETGs \label{dis:kinematics}}
A detailed study of stellar and gas kinematics in the CALIFA galaxy survey
is subject of several ongoing investigations within the CALIFA
collaboration \citep[][Falc\'on-Barroso et~al., in prep.]{GL13,SB14}, which
use higher-resolution (V1200; FWHM$\sim$2.3\AA) IFS data and to which the
reader is referred for detailed information. The brief remarks below are
merely meant to further highlight the kinematical diversity of our sample
ETGs.

As already revealed by several studies before, most notably in the the
SAURON and Atlas$^{\rm 3D}$ projects \citep[cf, e.g.,][]{Sarzi06,Sarzi10},
ETGs display a variety in their stellar kinematics patterns, from regular
rotation along the semi-major axis to nearly pressure-supported galaxies
with subtle, if any, signs of ordered motions.

Panels \rem{d}\&\rem{e} of Fig.~\ref{fig:bpt_n1167}--\ref{fig:bpt_n7550}
show that in most (27) of the ETGs studied here stellar kinematics are
dominated by rotation.  The velocity difference \dvstars=($v_{\rm a}+v_{\rm
  r}$)/2 between the receding ($v_{\rm r}$) and approaching ($v_{\rm a}$)
part of these systems is typically of the order of 200 \kmsec.
Interestingly, weak rotation at 60 \kmsec$\la$\dvstars$\la$ 110
\kmsec\ along the minor galaxy axis is clearly detected in three type~ii
ETGs (\object{NGC 6173}, \object{NGC 6338} and \object{NGC 7236}).  Five
galaxies (\object{UGC 10695}, \object{NGC 6515}, \object{UGC 11958},
\object{NGC 7550} and \object{UGC 29}) appear to be dominated by random
motions. Note, however, that the low spectral resolution and medium S/N of
our CALIFA V500 IFS data, do not permit to firmly rule out some degree of
rotation in these systems.  Higher-spectral resolution CALIFA data
(Falc\'on-Barroso et~al., in prep.)  enabling a detailed study of stellar
line-of-sight velocity patterns and possible kinematical distortions (e.g.,
counter-rotating cores) over the full set of CALIFA ETGs are necessary for
a conclusive assessment of this issue.  An interesting question in this
respect is how type~i/i+ and type~ii ETGs compare to each other with
respect to their angular momentum per unit mass in the definition by
\citet{Emsellem2007}. These authors have shown that fast rotators in the
SAURON sample tend to be lower-luminosity ETGs with well-aligned
photometric and kinematic axes, whereas slow rotators typically exhibit
significant misalignments and kinematically decoupled cores.

A comparative study of type i/i+ and ii ETGs with regard to the kinematics
of their \wim\ is obviously another key task that could help understanding
whether gas is internally produced, in the course of stellar evolution, or
accreted from captured satellites. The spiral-arm like SF features in
type~i+ ETGs, or faint embedded SF patterns in several other ETGs are
suggestive, yet no proof, for the gas accretion scenario. With regard to
the discussion below it is important to bear in mind that the non-detection
of ordered gas motions in most of our sample ETGs (especially those
classified type~ii) may reflect observational limitations due to the
extreme faintness of the \wim\ (see also Sect.~\ref{biases}, in particular
our remarks on the type i+ ETG \object{NGC 160} in
Fig.~\ref{fig:n0160}). Clearly, a detailed study of ionized-gas kinematics
is, in particular in type~ii ETGs a formidable task that only can be
pursued with a high S/N (typically $\geq$60 at 6390--6490 \AA) and/or
spatially binned data \citep[e.g.,][]{GL13}.

Some noteworthy trends are apparent though from the analysis of our
low-resolution V500 data: In roughly one third (11) of the sample ETGs (8
classified type~i and in the three type~i+) gas kinematics are consistent
with rotation aligned with stellar motions, albeit slightly perturbed
kinematical patterns. For instance, in \object{NGC 1167} the \wim\ rotation
axis appears slightly tilted relative to the stellar one, with a possible
spatial offset between the photometric and kinematical center.  In three
type~i ETGs (\object{NGC 5966}, \object{UGC 10695} and \object{UGC 10905})
gas kinematics appear to be dominated by outflows, and a suggestion for an
outflowing gas component is also present in \object{NGC 6146}.  Of the
three type~ii ETGs showing stellar rotation perpendicular to the major
axis, we do not detect gas motions in \object{NGC 6173} and \object{NGC
  7236}, whereas in \object{NGC 6338} there is a trace of a compact
rotating ($\sim$70 \kmsec) \wim\ component, centered on the optical
nucleus.  Likewise, in \object{NGC 7550} gas rotation at \dvgas$\sim 80$
\kmsec\ can be traced within the central $\sim$7~kpc of a
pressure-supported stellar host. Summarizing, this study further highlights
and adds observational insight into the kinematical diversity of the ETG
galaxy family \citep[see also, e.g.,][]{Sarzi06,Katkov2013}. On the other
hand, the evidence from Figs.~\ref{fig:bpt_n1167} and
\ref{fig:bpt_u05771}--\ref{fig:bpt_n7550} permits identification of broad
kinematical trends in the three ETG types: Type~i and i+ ETGs are generally
rotation-supported with many cases of perturbed rotational patterns or
outflows in their \wim, whereas type~ii ETGs exhibit both rotationally or
pressure supported stellar kinematics, and generally lack distinguishable
kinematic patterns in their ionized gas.  In Table~1 (col. 14) includes a
descriptive 3-character kinematical classification, with {\bf r} denoting
rotation-dominated kinematics, {\bf p} perturbed rotational patterns, {\bf
  o} gas outflows, and {\bf n} no distinguishable patterns or
pressure-supported stellar kinematics. The first and third character refer,
respectively, to the kinematics of stars and gas, and the second one
indicates in the case of stellar rotation (r or p) whether the kinematic
and photometric major axes are roughly aligned ($\parallel$) or not
($\not\parallel$).

\subsection{Equivalent width bias on ETG taxonomy\label{biases}}

\begin{figure*}
\begin{picture}(18.0,4.8)
\put(0.0,0.0){\includegraphics[height=4.9cm]{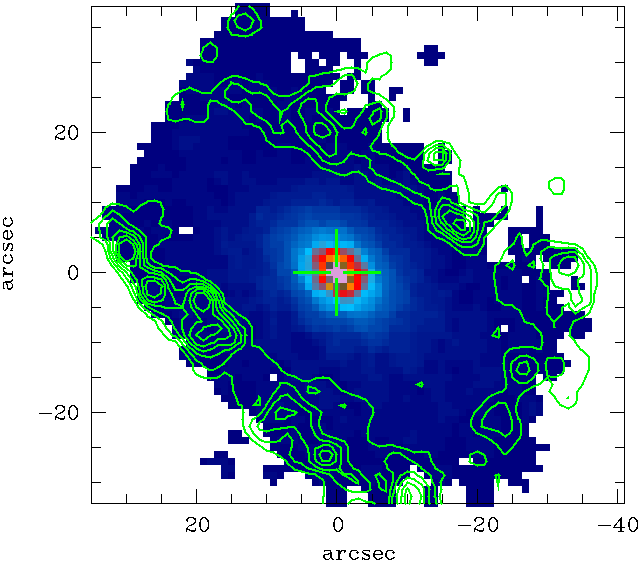}}
\put(5.8,0.0){\includegraphics[height=4.9cm]{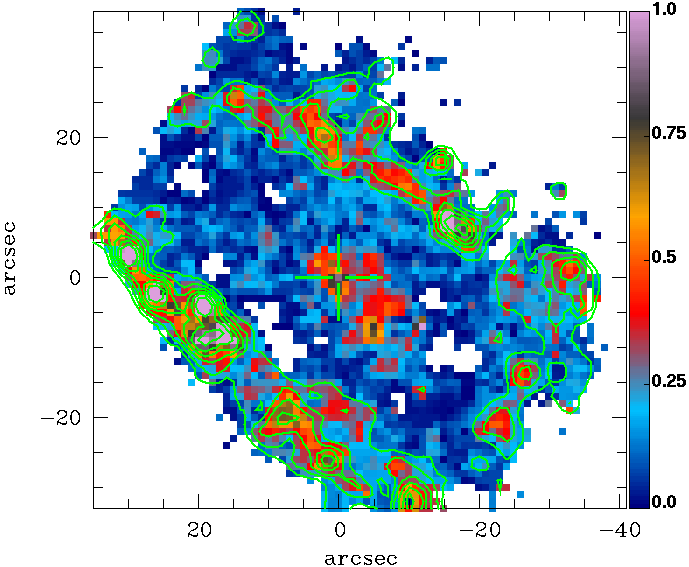}}
\put(12.0,0.0){\includegraphics[height=4.9cm]{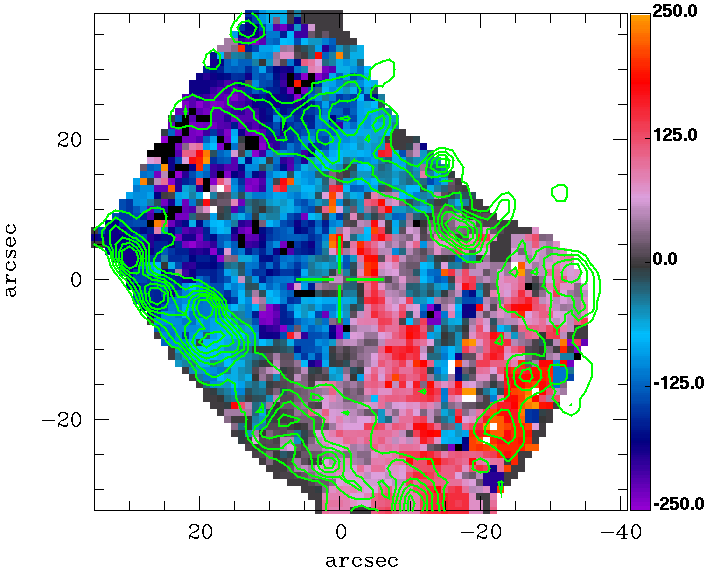}}
\PutLabel{1}{4.4}{\vcap a)}
\PutLabel{6.8}{4.4}{\vcap b)}
\PutLabel{13.0}{4.4}{\vcap c)}
\end{picture}

\caption[]{Emission-line free stellar continuum \rem{(a)}, \ha\ flux in
  $10^{-16}$ \uflux\ \rem{(b)} and \ha+[N{\sc ii}] line-of-sight velocity in
  km/s \rem{(c)} of the type~i+ ETG \object{NGC 160}, as inferred from
processing of CALIFA (v1.5) V500 with \p3d.  The overlaid contours delineate
the \ewha\ morphology of the galaxy and go from 4~\AA\ to 21~\AA\ in steps of
3~\AA. North is up and east to the left.}
\label{fig:n0160}
\end{figure*}

A central result from our study in P13 is the considerable radial
dependence of the \ewha\ in many ETGs (panels \rem{c}\&\rem{h}): A
distinctive property of several type~ii ETGs is their almost \wim-emission
free cores and their outwardly increasing \ewha, reaching in some cases
readily detectable levels ($\sim$\pAGBmean) in their periphery only. With
regard to type~i ETGs, although these show on average an
\ewha$\simeq$\pAGBmean\ in their extranuclear component, there are several
cases in our sample where the \ewha\ significantly varies within different
radial zones. This is particularly true for the two-zone \wim\ morphology
of type~i+ ETGs: An illustrative example \object{NGC 1349} showing a
constant \ewha$\simeq$\pAGBmean\ for \rr$\leq$8\arcsec\ (3.4 kpc), in
agreement with pure pAGB photoionization, whereas in the periphery
(\rr$\sim$16--20\arcsec) the \ewha\ rises to $\ga$6~\AA, revealing an extra
ionizing field produced by ongoing SF activity (Sect.~\ref{type-i}).

In fact, the intuitive expectation that emission-line EWs in ETGs are
spatially correlated with the surface brightness of the stellar background,
invariably attaining their peak value at the ETG nuclei, is disproved by
panels \trem{c} and \trem{h} of
Figs.~\ref{fig:bpt_n1167}--\ref{fig:bpt_n7550}. What a comparison of panels
\trem{a}\&\trem{b} with panel \trem{c} reveals in many cases is actually an
inverse trend with the \ewha\ showing -- contrary to the stellar and
\ha\ surface brightness -- low or even minimum value in the nuclear region
(e.g., Fig.~\ref{fig:bpt_n5966}). This spatial anti-correlation between
the intensity and EW of emission lines has been documented and discussed in
studies of low-mass starburst galaxies \citep{P98,P02} and, in the case of
ETGs, it can partly attributed to line-of-sight dilution of nuclear EWs by
a triaxial stellar host (P13). As we pointed out in P13, this geometric
effect introduces a selection bias against the detection of nuclear
activity in galaxy spheroids, since emission-line EWs (and obviously any
other manifestation of energy release by an AGN in UV-NIR wavelengths) in
their centers are diluted, not only by the local stellar background, but,
additionally, by the integrated line-of-sight luminosity of stars.  Since
this geometric dilution effect amplifies an observational bias stemming
from the high \lyc\ photon escape fraction in type~ii ETG nuclei, caution
is needed when arguing that the weakness of nuclear emission-lines and EWs
in ETGs/LINERs is evidence for weak or non-existent accretion-powered
nuclear activity.

In the light of these two effects, analogous considerations apply also to
the usage of \ewha\ as discriminator between AGN and `retired' galaxies in
the context of the WHAN classification \citep{cid11}: A nuclear
\ewha\ between \pAGBmin\ and \pAGBmax, i.e. in the range of predicted
values for pAGB photoionization in `retired' galaxies, does not necessarily
rule out AGN as the dominant excitation source of the \wim. The true
nuclear EW (i.e. the EW after correction for the geometric dilution and
\lyc\ escape bias) could exceed by more than one order of magnitude the
observed (projected) one, shifting on the WHAN diagram a seemingly `weak'
(low-luminosity) AGN into the `strong' AGN regime (cf our discussion in
P13).

It is also worth contemplating the observational biases that the EW
detection limit may have on taxonomical studies of ETGs. Note that
extranuclear nebular emission would not had been detected in any type~ii
ETG if our analysis was restricted to an \ewha$\geq$1~\AA.  Because of the
geometrical EW dilution effect, even the compact \ha-emitting nuclei of
many type~i and type~ii ETGs (e.g., in \object{NGC 1349}, \object{UGC
  8234}, \object{NGC 6515}, \object{NGC 7025}) vanish to EWs$\leq$1~\AA,
and would had gone undetected if such a `quality-ensuring'
\ewha\ threshold would had been adopted. By that cutoff, all but three
type~ii ETGs (\object{NGC 6338}, \object{UGC 11958} and \object{NGC 7550})
would had been classified as gas-free `passive' galaxies \citep[see
  definition by][]{cid11}.  Even though the thin higher-EW outer rim of
nebular emission would had been recovered in type~i+ ETGs, this alone would
not had demonstrated the extended nature of the circumnuclear
\wim\ component.

The importance of this EW-threshold bias may be illustrated through
inspection of the \wim\ morphology and kinematics maps of \object{NGC 160},
a type~i+ ETG which will be subject of a forthcoming article of this
series.  The overlaid contours in the three panels of Fig.~\ref{fig:n0160}
depict the \ewha\ morphology of the galaxy and go from 4~\AA\ to 21~\AA\ in
steps of 3~\AA. Whereas a substrate of diffuse (0.5$\la$\ewha [\AA]$\la$2)
\wim\ is present almost everywhere in the central part of the galaxy, the
bulk of \ha\ emission is confined to an extended
($\sim$60\arcsec$\times$40\arcsec) rim surrounding the low-surface
brightness stellar periphery \citep[see also][]{GL13}.  An EW detection
threshold of 4~\AA\ (i.e. $\geq$\pAGBmax; outermost contour in all panels)
would obviously allow for the recovery of the outer rim only, leading to
the erroneous conclusion that this ETG is completely devoid of \wim\ almost
throughout its optical extent, out to $\sim$24 \sbb\ in the SDSS $r$ band
(cf Breda et al., in prep.).  A further implication would had been the
complete rejection of the pAGB photoionization hypothesis, since this is
testable only for an EW detection threshold $\leq$\pAGBmax, and the
conclusion that photoinization by OB stars is the sole gas excitation
mechanism in this ETG.  An EW threshold $\ga$\pAGBmax\ would also result in
a severely incomplete view of the overall gas kinematics everywhere but in
the EW-enhanced rim, with obvious implications to the dynamical modeling of
such a galaxy.

\section{Summary and conclusions \label{summary}}
In this article, we investigate in significantly more detail the sample of
32 CALIFA ETGs that we discussed in P13 which had its focus on their radial
\ewha\ distribution and \lyc\ photon escape fraction.  We study here the 2D
properties of their diffuse warm interstellar medium (\wim) to gain insight
into its excitation mechanism(s) and provide additional observational
constraints aiming at systematizing the physical properties of these
systems.  Among other quantities derived by applying our IFS data
processing pipeline \p3d\ to low-spectral resolution ($\sim$6~\AA\ FWHM)
CALIFA data, we present for each galaxy \ha\ intensity and EW maps, as well
as radial profiles. Also, we analyze the nuclear and extranuclear nebular
emission of the \wim\ using diagnostic line ratios.  A comparison sample of
intermediate-to-late type CALIFA galaxies is meant to illustrate how
various observables of interest change from bulge-dominated ETGs towards
disk-dominated star-forming (SF) galaxies.  Additionally, we briefly
discuss the kinematical properties of gas and stars in our sample.

The main conclusions from this study may be summarized as follows:

\trem{i)} The analyzed ETGs contain an extended \wim\ component displaying
a considerable diversity in its mean \ewha\ and its radial gradients.  As
we discussed in P13, our ETG sample can be tentatively subdivided in two
groups. The extranuclear \wim\ of the first class (type~i) is characterized
by a nearly constant \ewha\ of $\sim$1~\AA, with the exception of a few
systems \citep[i+, in the notation by][]{G14a} showing a steep
\ewha\ increase in their periphery.  The second group (type~ii) displays
almost \wim-evacuated cores and a gradual increase of the \ewha\ outwardly
to $\la$1~\AA.

\trem{ii)} The ETGs studied here also display a large diversity in their
kinematics. The majority (27 out of 32) of our sample ETGs exhibit stellar
kinematics dominated by rotation, with three systems showing a kinematic
major axis aligned with the photometric minor axis.  Type~i and i+ ETGs are
generally rotation-supported with many cases of perturbed rotational
patterns or outflows in their ionized gas, whereas the class of type~ii
ETGs contains both rotationally and pressure supported stellar systems,
which generally lack distinguishable kinematic patterns in their ionized
gas.

\trem{iii)} We compute the time evolution of the \ewha\ for various star
formation histories (SFHs) and stellar metallicities based on an
evolutionary synthesis model and assuming case~B recombination and standard
conditions for the gas. We find that SFHs involving an exponentially
decreasing star formation rate since 13 Gyr can reproduce the low
($\sim$1~\AA) \ewha's in the extranuclear component of present-day ETGs
when a short ($\leq$1~Gyr) e-folding timescale is adopted. In this case,
the Lyman continuum (\lyc) output predicted to arise from the pAGB stellar
component is sufficient for powering the excitation of the diffuse \wim.
SFHs involving a longer e-folding timescale (e.g., $\geq$3 Gyr) imply
higher \ewha's than observed ones.

\trem{iv)} As far as most type~i ETGs are concerned, photoionization by the
evolved pAGB stellar background offers a consistent, yet not necessarily
compelling, explanation for the observed \ewha's and $\tau$ ratios (the
inverse ratio of the observed \ha\ luminosity to that predicted for pure
pAGB photoionization).  We point out that \lyc\ photon escape and geometric
line-of-sight dilution of EWs (cf P13) could conspire in such a way as to
reproduce the low ($\simeq$1~\AA) and nearly constant \ewha\ in the
extranuclear component of type~i ETGs, thereby mimicking a predominance of
pAGB photoionization to the excitation of their diffuse ionized gas
component, while hiding a potentially important role by AGN and low-level
SF.  The presence of the latter in the extranuclear component of several
ETGs from our sample is clearly indicated from \ewha\ maps.

\trem{v)} The radial \ewha\ gradients and the extreme faintness of nebular
emission in ETGs, in particular in systems classified type~ii, leads to
biases impacting the detectability of the extended \wim\ component and the
interpretation of its excitation mechanisms. We point out that nebular
emission in the centers of type~ii ETGs will generally evade detection
below an EW threshold of $\sim$0.5--1.0~\AA, which could lead to the
conclusion that these systems are entirely devoid of \wim, and their
erroneous classification as passive galaxies.

\trem{vi)} The radial distribution of diagnostic line ratios indicates
that, in their majority, ETGs show LINER-specific spectroscopic properties
both in their nuclear and extranuclear zones. Our sample includes, however,
a few notable exceptions (in particular, type~i+ ETGs) where a significant
contribution by star formation in peripheral zones is reflected both on
radial \ln2ha, \lo3hb\ and $\tau$ profiles, and \ewha\ maps. This calls
attention to the heterogeneity of the ETG class, pointing to a complex
interplay between the 3D properties of the \wim\ and the combined action of
various gas ionization sources, such as, e.g., shocks, OB stars and the
evolved pAGB component. As we argue in \citet{P13} and further detail here,
the weakness of nebular emission lines in ETG nuclei is not compelling
evidence for the absence of accretion-powered nuclear activity in these
systems. In fact, radio-continuum and/or X-ray data, or bipolar or unipolar
\wim lobes protruding several kpc away from the LINER nuclei of 25\% of our
sample ETGs witness the energetic action of an AGN.  Summarizing, the
results and consideration presented in this study are incompatible with the
hypothesis of the global gas excitation patterns in ETGs being controlled
by one single dominant mechanism.

\begin{acknowledgements}
  This paper is based on data from the Calar Alto Legacy Integral Field
  Area Survey, CALIFA (http://califa.caha.es), funded by the Spanish
  Ministery of Science under grant ICTS-2009-10, and the Centro
  Astron\'omico Hispano-Alem\'an. JMG acknowledges support by
  Funda\c{c}\~{a}o para a Ci\^{e}ncia e a Tecnologia (FCT) through the
  Fellowship SFRH/BPD/66958/2009 and POPH/FSE (EC) by FEDER funding through
  the program Programa Operacional de Factores de Competitividade
  (COMPETE). PP is supported by FCT through the Investigador FCT Contract
  No. IF/01220/2013 and POPH/FSE (EC) by FEDER funding through the program
  COMPETE. JMG\&PP also acknowledge support by FCT under project
  FCOMP-01-0124-FEDER-029170 (Reference FCT PTDC/FIS-AST/3214/2012), funded
  by FCT-MEC (PIDDAC) and FEDER (COMPETE). SFS acknowledges support from
  CONACyT-180125 and PAPIIT-IA100815 grants. Support for LG is provided by
  the Ministry of Economy, Development, and Tourism's Millennium Science
  Initiative through grant IC120009, awarded to The Millennium Institute of
  Astrophysics, MAS. LG acknowledges support by CONICYT through FONDECYT
  grant 3140566. CJW acknowledges support through the Marie Curie Career
  Integration Grant 303912. RGB is supported by the Spanish Ministerio de
  Ciencia e Innovaci\'on under grant AYA2010-15081. R.A. Marino is funded
  by the Spanish program of International Campus of Excellence Moncloa
  (CEI). IM acknowledges financial support by the Junta de Andaluc\'ia
  through project TIC114, and the Spanish Ministry of Economy and
  Competitiveness (MINECO) through projects AYA2010-15169 and
  AYA2013-42227-P. AdO acknowledges financial support from the Spanish
  grants AYA2010-15169 and AYA2013-42227-P.  J. F-B acknowledges support
  from grant AYA2013-48226-C3-1-P from the Spanish Ministry of Economy and
  Competitiveness (MINECO). The {\sc starlight} project is supported by
  the Brazilian agencies CNPq, CAPES, and FAPESP.  We benefited from
  stimulating discussions with several members of the CALIFA collaboration.
  This research made use of the NASA/IPAC Extragalactic Database (NED)
  which is operated by the Jet Propulsion Laboratory, California Institute
  of Technology, under contract with the National Aeronautics and Space
  Administration.
\end{acknowledgements}



\appendix
\section{The time evolution of the H$\alpha$ equivalent width
in the case of pure pAGB photoionization \label{pAGB}}
As discussed in Sect.~\ref{intro}, the time evolution of the \lyc\ output
from the evolved pAGB stellar component in ETGs has been evaluated in
several previous studies \citep[e.g.,][]{bin94,sta08,cid10} all of which
assume an instantaneous SF process. Here we extend this theoretical work by
addressing how the \lyc\ rate from the pAGB stellar component, and the
associated nebular emission by photoionized gas, evolve when a slowly declining SF
activity is assumed. For this purpose, we use our evolutionary synthesis
code, adopting a wider range of star formation histories (SFHs), from an
instantaneous burst to continuous SF to compute synthetic spectral energy
distributions (SEDs).

The synthetic composite stellar SEDs were computed using the full set of
ages for the SSPs from \citet[][hereafter BC03; comprising 221 spectra
  spanning an age between 0 and 20 Gyr]{bru03} and assuming a constant
metallicity Z$_\odot$. The BC03 SSP library uses `Padova 1994' evolutionary
tracks \citep{alongi,bressan,fagottoa,fagottob,fagottoc,girardi}, the
\citet{chabrier} IMF with a lower and upper mass cutoff of $0.1$ and 100
M$_{\odot}$, respectively, and the STELIB stellar library
\citep{LeBorgne2003}, supplemented blueward of 3200 \AA\ and redward 9500
\AA\ by BaSel $3.1$ spectra \citep{Westera2002}. The UV coverage of these
SSPs permits computation of the \lyc\ photon output as a function of time
(see Sect.~\ref{P3D-m3}).  In computing the nebular continuum contribution
we assumed an electron temperature and density of $T_{\rm e} = 10^4$~K and
$n_e = 100$ cm$^{-3}$, respectively. Balmer emission-lines were computed
from the total UV ionizing flux using the effective recombination
coefficient $\alpha^{\rm eff}_{H\alpha}$ and by assuming ionization-bound
nebulae (case~B recombination).
  
Figure~\ref{fig:rebetiko} (upper panel) shows the time evolution of the
\ewha\ for four parametrizations of the SFH: instantaneous burst,
continuous SF and exponentially decreasing SF with an e-folding time of 1
Gyr and 3 Gyr (black, yellow, green and violet curve, respectively). The
\ewha\ expected from the pAGB ($\geq 10^8$ yr) stellar component for the
combination of these models is displayed as the cyan shaded area.  The
lower-left panel of the figure shows the time evolution of the \lyc\ photon
rate per per solar mass, with the fractional contribution of the pAGB
component displayed separately in the lower-right diagram.

\begin{figure}
  \begin{picture}(8.6,8.6)
    \put(-0.5,0.0){\includegraphics[width=9.5cm,height=10cm, clip=true,
      viewport=0 180 580 735]{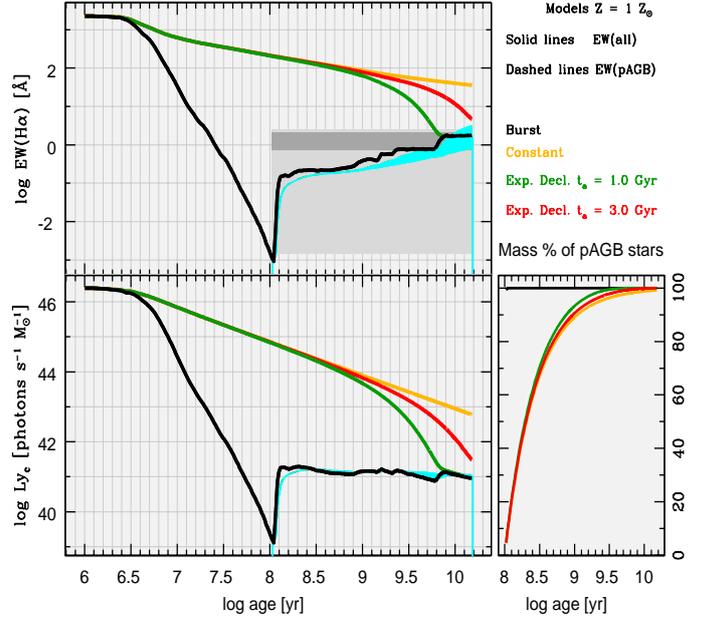}}
\end{picture}

\caption[]{\rem{upper panel:} Predicted \ewha\ (\AA) as a function of time
  for a solar-metallicity stellar population with a Salpeter IMF forming
  instantaneously (black), continuously at a constant SFR (yellow) and with
  an exponentially decreasing SFR with e-folding times $t_{\rm e}$ of 1~Gyr
  and 3~Gyr (green and red curve, respectively).  The light-grey shaded
  area in the lower-right part of the diagram encompasses the range in
  \ewha\ predicted from pAGB photoionization for ages $\geq$0.1 Gyr, and
  the narrow dark strip depicts the corresponding prediction
  (\ewha$\simeq$1 \AA\ $\equiv$\pAGBmean) for the typical age of ETGs
  ($\geq$6 Gyr). The \rem{lower panel} shows the time evolution of the
  \lyc\ photon rate expected by the instantaneous and exponentially
  decreasing SFR models, and the mass fraction in \% of stars older than
  100 Myr (referred to as pAGB) is displayed separately in the lower-right
  diagram.}
\label{fig:rebetiko}
\end{figure}

From Fig.~\ref{fig:rebetiko} it is apparent the well-known fact \citep[see,
  e.g.,][]{Starburst99,Weilbacher01,Kotulla09-GALEV} that the \lyc\ photon
rate and the corresponding Balmer-line EWs show a strong evolution with
time, depending on the assumed SFH.  For an instantaneous burst model, the
\lyc\ photon rate (s$^{-1}$), following a steep decrease by $\sim$5~dex
over the first $10^8$ yr of galaxy evolution, levels off to a nearly
constant level, scaling only with the available stellar mass as $\approx$
$10^{41}$ \mstar/\msun\ \citep[see also][]{cid11}.  This results in a
nearly constant \ewha\ in the narrow range between $\sim$0.1~\AA\ and
$\sim$2.4~\AA\ (\pAGBmax) over nearly the whole Hubble time.  Note that an
\ewha$\simeq$0.1 \AA\ corresponds to a short ($<$50 Myr) period at the
onset of the pAGB phase.  As the curves within the shaded area of
Fig.~\ref{fig:rebetiko} depict, the \ewha\ predicted for the pure pAGB
component for other SFHs also shows little evolution with time, with the
notable trend, however, of remaining lower than that predicted for the
instantaneous burst model. This is particularly true for the continuous SF
model, which implies for the \ewha\ owing to the pAGB component alone a by
a factor $\sim$2 lower value than the single-burst model.  However, by
considering the \ewha\ expected to arise from the total (young$+$pAGB)
stellar component (curves in the upper part of the same panel), it is
obvious that continuous as well as exponentially decreasing ($t_{\rm e}$=3
Gyr) SF models can be discarded for ETGs, as they predict a large
($\geq$6~\AA) \ewha\ even for an age of $\sim$10~Gyr, whereas the
observations (cf, e.g., P13 and Sect.~\ref{results}) indicate values in the
range of 1--2~\AA. On the other hand, a model involving a shorter ($t_{\rm
  e}$=1 Gyr) phase of exponentially declining SFR (green curve) is
compatible with the literature data, as it predicts an
\ewha$\approx$\pAGBmax\ after 6.3 Gyr of galactic evolution, just like the
instantaneous SF model. For reasonable assumptions on the age of ETGs
($\geq$6~Gyr) these last two SFHs imply consistently a narrow span between
$\sim$1~\AA\ (\pAGBmean) and \pAGBmax\ for the \ewha\ of photoionized gas.
Allowing for a generous low limit of \pAGBmin=0.5~\AA\ (corresponding to a
minimum $\tau$=2 defined in P13 as a conservative low limit for the
presence of \lyc\ escape) we consider an observed \ewha\ in the range
between 0.5~\AA\ and 2.4~\AA\ to be consistent with the pAGB
photoionization hypothesis.

\section{Analysis methodology \label{meth}}

Figure~\ref{fig:Porto3D} provides a schematic overview of the three main
modules of \p3d\ (v.2), which are briefly described next.

\begin{figure*}[t]
\includegraphics[width=11cm, viewport=-40 00 520 590]{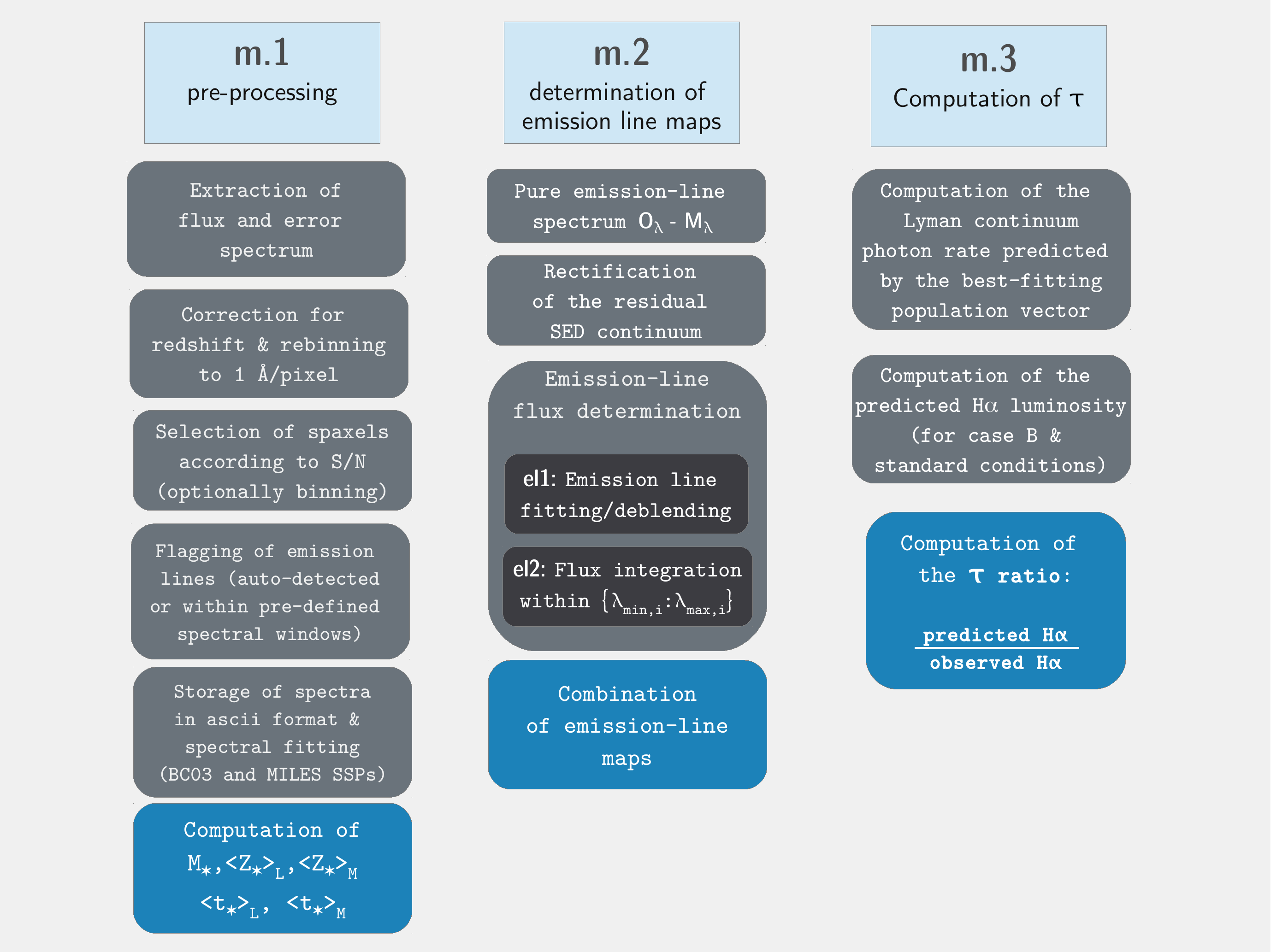}

\caption[]{Schematic outline of the three main modules of \p3d\ (v.2).}
\label{fig:Porto3D}
\end{figure*}

\subsection{Module m.1: Computation of stellar fits and post-processing of the results \label{P3D-m1}}
The module \rem{m.1}, which is invoked at the first stage, extracts spectra
from an IFS cube spaxel-by-spaxel, along with auxiliary information
provided by the CALIFA data reduction pipeline (error spectrum, bad pixel
map), computes initial statistics, and performs a number of preparatory
steps that are necessary for subsequent fitting with \SL. These include the
computation of the signal-to-noise (S/N) in various spectral windows (with
the option of rejecting spaxels that do not satisfy certain quality
criteria), the flagging of spurious features and emission lines within
pre-defined or automatically generated spectral masks, and the spectral
resampling to the restframe and wavelength rebinning to 1~\AA/pixel.  The
observed flux at each wavelength $O_{\lambda}$ and its uncertainty
$e_{\lambda}$, provided by the CALIFA v1.3c reduction pipeline output or
auto-computed with a sliding boxcar filter\footnote{This approach has been
  adopted in our pilot study of the ETGs \object{NGC 5966} and \object{NGC
    6762} in K12, as an error spectrum was not provided in earlier versions
  of the CALIFA data reduction pipeline.}, are finally exported together
with flagging tags into ascii files in the format required for subsequent
fitting with \SL.  This module also creates up to 16 execution scripts,
facilitating automated parallel spectral fitting of the data with
multi-core CPUs.  For each sample galaxy, typically $\sim$2700 individual
spectra with a S/N$\ga$30 at the normalization wavelength (5150 \AA) were
extracted and modeled with \SL.

The goal of \SL\ is the decomposition of an observed spectrum into a linear
combination of its constituent SSP spectra.  A description of the code is
given in Cid Fernandes et al. 2005 (C05), 2007; and subsequent publications
of the SEAGal collaboration\footnote{STARLIGHT \& SEAGal:
  http://www.starlight.ufsc.br/ } and is supplemented by a detailed
cookbook for its application.

The best-fitting set of SSPs, together with the derived intrinsic $V$ band
extinction and velocity dispersion of the stellar component are referred to
in the following as population vector. This encapsulates all information of
the best-fitting model spectrum $M_\lambda$
\begin{center}
$M_\lambda =M_{\lambda_0} \left[ \sum\limits_{j=1}^{N_\star}
    x_j~b_{j,\lambda}~r_\lambda \right] \otimes G(v_\star,\sigma_\star)$,
 \end{center}
where $b_{j,\lambda} \equiv L_\lambda^{SSP}(t_j,Z_j)
/L_{\lambda_0}^{SSP}(t_j,Z_j)$ is the spectrum of the $j^{\rm th}$ SSP
normalized at $\lambda_0$, $r_\lambda \equiv 10^{-0.4(A_\lambda -
  A_{\lambda_0})}$ is the reddening term, {\boldmath $x$} is the population
vector, $M_{\lambda_0}$ is the synthetic flux at the normalization
wavelength, $N_\star$ is the total number of SSPs,
$G(v_\star,\sigma_\star)$ is the line-of-sight stellar velocity
distribution, modeled as a Gaussian centered at velocity $v_\star$ and
broadened by $\sigma_\star$.

The convergence of the solution is driven by the standard $\chi^2$
minimization with respect to the observed spectrum $O_{\lambda}$
\begin{center}
$\chi^2(x,M_{\lambda_0},A_V,v_\star,\sigma_\star) =
  \sum\limits_{\lambda=1}^{N_\lambda} \left[ \left(O_\lambda - M_\lambda
    \right) w_\lambda \right]^2$,
\end{center}
whereby spectral weights $w_\lambda$ are given by the inverse of the error
spectrum $e_{\lambda}$.

\SL\ fits were computed in the spectral region between 4000 \AA\ and 6800
\AA. This was mainly because the S/N of our sample ETGs bluewards of 4000
\AA\ is generally too low for a reliable stellar fit and the subsequent
determination of the flux in the [O{\sc ii}]$\lambda\lambda$3726,3729 line
doublet. The red spectral region (6900--7300 \AA) was also disregarded
because of its generally low S/N and spectral artifacts due to vignetting
in the outer part of the PPAK FOV \citep[see][]{Sanchez2012}.

For each spaxel \SL\ fits employing between 7 and 12 Markov chains were
computed twice, using two different Simple Stellar Population (SSP)
libraries, each comprising 34 ages (between 5 Myr and 13 Gyr) and three
metallicities (0.008, 0.019, 0.03), i.e. 102 SSPs in total.  The first set
of models uses SSPs from \citet[][hereafter BC03]{bru03}.  As these cover
the UV spectral range, they allow to infer the \lyc\ photon output from
each best-fitting population vector (see Sect.~\ref{pAGB} for further
details) which is in turn used for the computation of the $\tau$ ratio
(module \rem{m.3}). For the second set of \SL\ fits we used SSPs from MILES
\citep{san06,vaz10,FB11} which cover the optical spectral range (3540 --
7410 \AA) only.

The module \rem{m.1} also permits spaxel binning prior to fitting according to
various prescriptions (e.g., averaging to a constant S/N along azimuthal
segments within isophotal annuli, see discussion in
Sect.~\ref{radial_profiles} and Fig.~\ref{fig:segmentation}). Whereas a number
of parallel runs on binned data cubes were conducted to check the robustness
of the detection of faint nebular emission in the extranuclear component of
our sample ETGs, we resorted to the methodology in K12 and P13 of carrying the
spectral analysis in a single-spaxel (\sisp) mode, and subsequently combining
the results within irregular annuli (Sect.~\ref{radial_profiles}).

Figure~\ref{fig:SL_fit} shows a high-quality \citep[mean percentual deviation 
  ADEV$\simeq$1.4; cf \SL\ manual and][]{cid05} fit to a high-S/N spectrum
from an off-center spaxel in the ETG \object{NGC 5966}.  The observed
spectrum $O_{\lambda}$, normalized at 5150 \AA, and the fit ($M_{\lambda}$) to
it are shown in orange and blue color, respectively, in panel \rem{a}.  The
lower-left panel \rem{b} displays the fit residuals, with shaded regions
depicting spectral regions flagged prior to the fit, or subsequently rejected
by \SL.  Panels \rem{c}\&\rem{d} show, respectively, the luminosity ($x_j$)
and mass ($\mu_j$) contribution of the individual SSPs (1 \dots $j$) composing
the best-fitting population vector. The thin grey vertical lines in the
upper-right panel mark the 34 ages available in the SSP library for the three
stellar metallicities (each coded by a different color), and the shaded area
in the lower panel shows a smoothed version of the $\mu_j$ distribution.  The
latter represents an illustrative approximation to the true SFH of stellar
populations within the spaxel under study.  It can be seen that the SFH
involves an extended star formation episode occuring until $\sim$7 Gyr ago 
with a recent SF peak at $\sim$10 Myr.
This fit therefore indicates that the stellar mass \mstar\ is dominated by an
evolved pAGB stellar population with a mass-weighted age of 8.2 Gyr. It
also implies little intrinsic extinction ($\la$0.2 $V$ mag) and no significant
star-forming activity over the past $\sim$1~Gyr.

\begin{figure*}
\begin{picture}(18.4,8.2)
\put(0,-4.4){\includegraphics[width=0.64\textwidth, viewport=20 30 520 290]{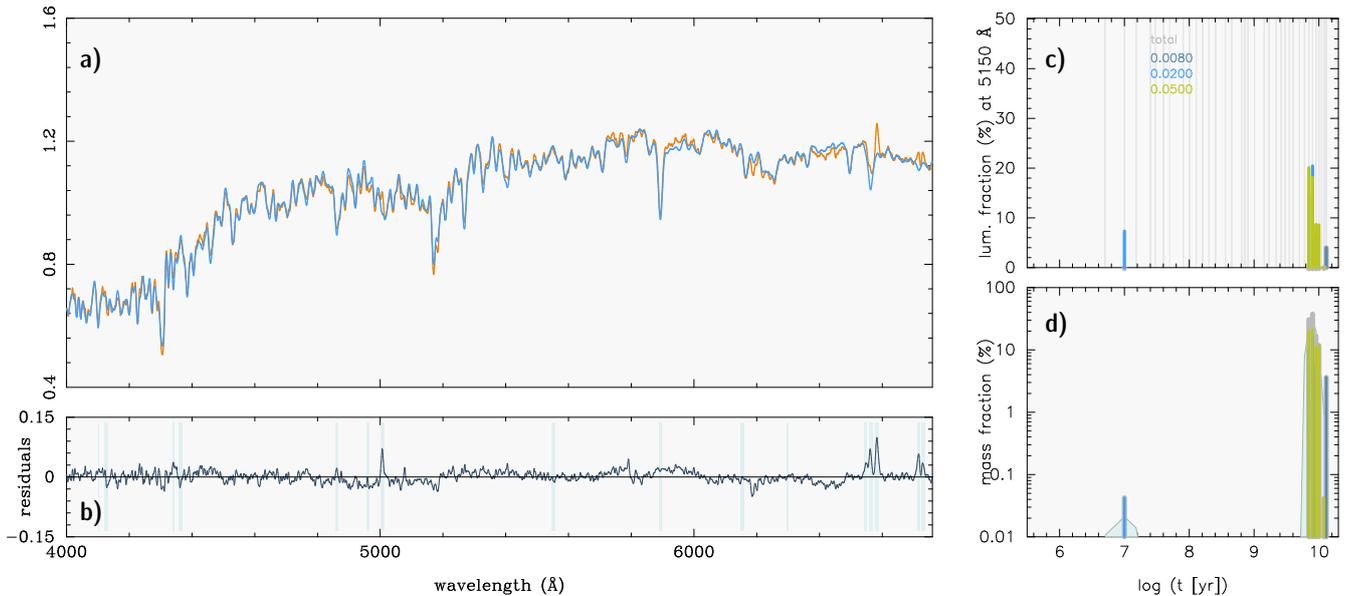}}
\PutLabel{1.2}{7}{\vcap a)}
\PutLabel{1.2}{1}{\vcap b)}
\PutLabel{13.9}{7}{\vcap c)}
\PutLabel{13.9}{3.5}{\vcap d)}
\end{picture}
\caption[]{\rem{a:} Example of a high-quality fit with \SL\ to a V500 spectrum
  (blue and orange color, respectively) extracted from an off-nucleus spaxel
  of the ETG \object{NGC 5966}.  From the lower-left panel \rem{b} it is apparent that
  the fit residuals are small (on average, $\la$5\% of the stellar continuum
  level) yet systematic (see also Fig.~\ref{fig:SL_residuals} and discussion
  in Sect. \ref{P3D-m2}). Shaded areas depict wavelength intervals that were flagged 
  prior to spectral fitting. Panels \rem{c} and \rem{d} display, respectively,
  the luminosity ($x_j$) and mass ($\mu_j$) contribution of individual SSPs (1
  \dots $j$) in the best-fitting population vector. The thin grey vertical
  lines in the upper-right panel depict the 34 ages available in the SSP
  library for three stellar metallicities (color coded), and the shaded area
  in the lower panel shows a smoothed version of the $\mu_j$, which is meant
  as an illustration of the star formation history. The fit implies a dominant
  old (pAGB) stellar population suffering little intrinsic extinction ($A_V\la
  0.2$ mag).}
\label{fig:SL_fit}
\end{figure*}

The module \rem{m.1}, besides storing the relevant output from each fit
($\sigma_{\star}$, $v_{\star}$, ..., among others), computes various
quantities of interest, such as the time at which 50\% of the present-day
and ever-formed \mstar\ was in place, the luminosity contribution of stars
formed in the past 5~Gyr (\m5), and the luminosity- and mass-weighted mean
stellar age ($\left<t_\star \right>_L$ and $\left<t_\star \right>_M$,
respectively) as
\begin{equation}
\left<\log\, t_\star \right>_L = \sum_{j=1}^{N_\star} x_j\, \log\, t_j,
\end{equation}
and 
\begin{equation}
\left<\log\, t_\star \right>_M = \sum_{j=1}^{N_\star} \mu_j\, \log\, t_j.
\end{equation}
The dispersion of these quantities is also computed in the standard manner
and exported into a data cube with dimension $x \times y \times z$, where $x \times y$ is the dimension 
of the CALIFA input IFS cubes in arcsec, and $z$ holds various quantities 
(age, metallicity, kinematics, emission-line fluxes, etc.) for each spaxel.

\begin{figure}
\begin{picture}(8.6,11.0)
\put(2,5.8){\includegraphics[width=6.0cm, viewport=20 20 520 270]{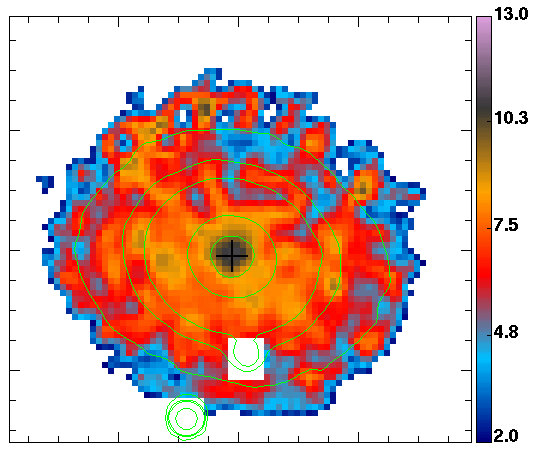}}
\put(2,0.3){\includegraphics[width=6.0cm, viewport=20 20 520 270]{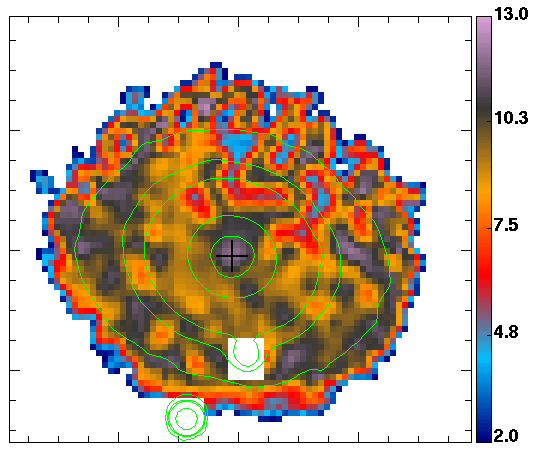}}

\PutLabel{+4.57}{-0.03}{\yx 0}
\PutLabel{+5.95}{-0.03}{\yx 20}
\PutLabel{+3.00}{-0.03}{\yx -20}
\PutLabel{+4.35}{-0.23}{\yx arcsec}
\PutLabel{1.65}{2.50}{\rotatebox{90}{\yx 0}}
\PutLabel{1.45}{2.30}{\rotatebox{90}{\yx arcsec}}
\PutLabel{1.65}{0.90}{\rotatebox{90}{\yx -20}}
\PutLabel{1.65}{3.90}{\rotatebox{90}{\yx 20}}

\PutLabel{1.65}{8.00}{\rotatebox{90}{\yx 0}}
\PutLabel{1.45}{7.80}{\rotatebox{90}{\yx arcsec}}
\PutLabel{1.65}{6.40}{\rotatebox{90}{\yx -20}}
\PutLabel{1.65}{9.40}{\rotatebox{90}{\yx 20}}

\PutLabel{+8.20}{1.20}{\rotatebox{90}{\color{greyJMG}\yx Mass-weighted stellar age [Gyr]\color{black}}}
\PutLabel{+8.20}{6.60}{\rotatebox{90}{\color{greyJMG}\yx Luminosity-weighted stellar age [Gyr]\color{black}}}
\end{picture}
\caption[]{Luminosity-weighted and mass-weighted stellar age (upper and
  lower map, respectively) in Gyr (right-hand vertical bar) determined from
  \SL\ fits to individual spaxels. A trend for decreasing age with radius
  is apparent, in qualitative agreement with the results by
  \citet{Rosa14a}. The \tmass\ map indicates a mass-weighted stellar age of
  $\geq$7 Gyr all over the optical extent of the ETG. }
\label{fig:tmass}
\end{figure}

The \tmass\ and \m5\ maps provided by \p3d\ (see Fig.~\ref{fig:tmass} for
an example) were only used for checking the soundness of the \SL\ fits and
checking the pAGB nature of the stellar component in the studied ETGs.

\subsection{Module m.2: Computation of emission-line maps \label{P3D-m2}}
The extraction of emission line fluxes in ETGs is in itself a challenging task due to their faintness, as already pointed out in 
several previous studies \citep[e.g.,][, K2, P13]{Sarzi06,ani10}. Indeed, faint (EW$\sim$1~\AA) emission lines, buried within broader 
stellar absorption features, can only be detected and quantitatively studied upon precise subtraction of a high-S/N stellar continuum. 
Even in that case, a careful assessment of their uncertainties in individual spaxels is fundamental.

Previous longslit and IFS studies have generally investigated the diffuse
\wim\ component of ETGs down to EW levels of $\ga$2~\AA, i.e. of the order
of the EWs of Balmer stellar absorption profiles, leading to the conclusion
that 2/3 of ETGs show faint nuclear and extranuclear nebular emission.
A key contribution to advancing our understanding of the
\wim\ component in ETGs has been made by the SAURON project, which has first
disclosed the diversity of emission-line morphologies and kinematics in these systems. 
Quite importantly, \cite{Sarzi06} reported the detection of very faint
(EW$\sim$0.1~\AA) \hb\ emission in some ETGs, showing that even systems 
that were previously thought to be entirely devoid of ionized gas actually
contain a low-surface brightness substrate of \wim. 

It is important to bear in mind, on the other hand, that emission-line flux
determinations in ETGs are sensitively dependent on the quality of stellar
fits, which in many cases show subtle yet systematic residuals.  An example
is given in Fig.~\ref{fig:SL_residuals} where we show a typical,
medium-quality fit to a \sisp\ high-S/N spectrum.  From the upper panel it
can be seen that fitting residuals are minor (on average, $\leq$5\% of the
normalized continuum), yet systematic with both narrow ($\sim$40 \AA) and
broader ($\sim$200~\AA) spectral intervals imperfectly fitted.  These
residuals were discussed in \citep[e.g.,][among others]{Sarzi06,Walcher06}
and ascribed to a combination of various causes (e.g., imperfections in
SSPs and/or fitting codes, $\alpha$ element enhancement, contamination from
telluric line emission). In the framework of CALIFA this issue was further
analyzed in \cite{Husemann2013} through median-averaging of several
thousands of individual spectra, showing indeed a systematic behavior in
the residuals, in agreement with a parallel study by our team. The latter
has shown that broad ripples in fitting residuals are partly to
inaccuracies in the spectral sensitivity function adopted in the CALIFA
v1.3c reduction pipeline. Indeed, a reprocessing with \p3d\ of part of the
IFS data after reduction with the upgraded version v1.4 shows that these
broad residuals are much reduced or absent, although smaller-scale
residuals generally persist.  This issue is currently under scrutiny within
the CALIFA collaboration.

On the other hand, as pointed out in \cite{Husemann2013}, since these
residuals are of the order or lower than the continuum noise level, their
impact on emission line determinations is in practice negligible.  Whereas
we concur with this statement, based on the analysis of a few emission-line
galaxies in our comparison sample (see Appendix \ref{LTGsample}), we find
that small-scale fitting residuals can significantly impact flux
determinations for faint (EW$\sim$1~\AA) emission lines when the standard
technique of approximating the local rest-continuum by a first-order
polynomial is used.  In this case, Gaussian line fitting/deblending may
fail to converge, or could lead to large errors owing to small-scale
variations of the local background beneath and adjacent to emission lines.
For this reason, these systematic fitting residuals could effectively place
an EW limit of $>$2~\AA\ on studies of the \wim, even when high-S/N spectra
are available.

\begin{figure*}
\begin{picture}(18.4,7.4)
\put(0.0,4.6){\includegraphics[width=18.0cm, clip=true, viewport=90 100 554 180]{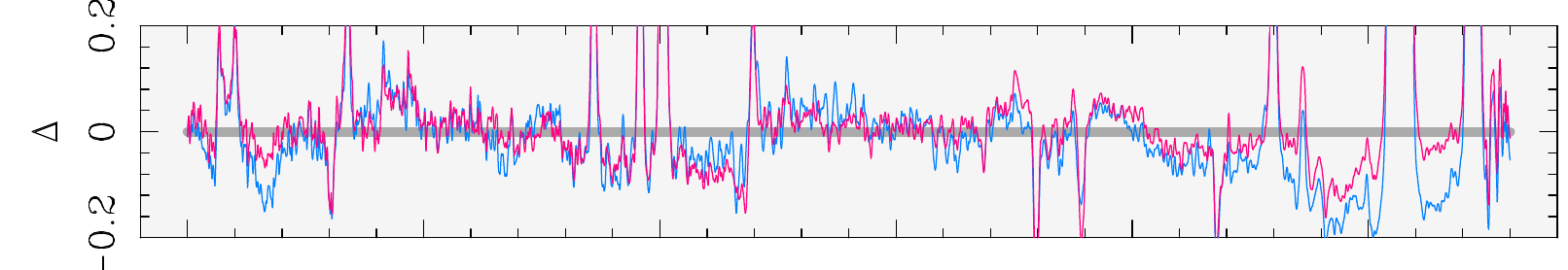}}
\put(0.0,2.4){\includegraphics[width=18.0cm, viewport=90 130 554 180]{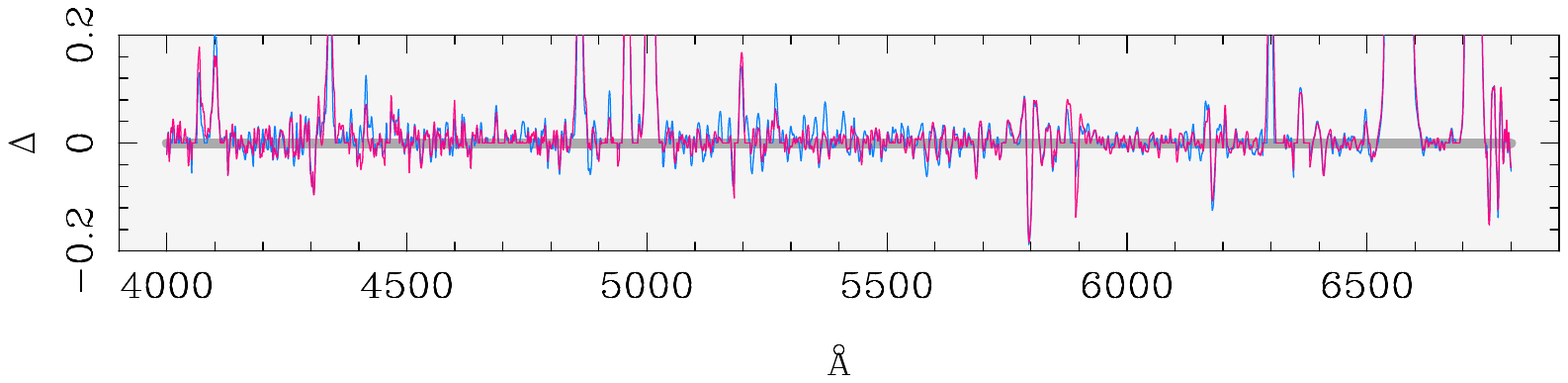}}
\end{picture}

\caption[]{\rem{upper panel:} Subtraction of the best-fitting stellar SED computed with \SL\ with two different SSP libraries (BC03 and MILES; blue and 
magenta color, respectively) from an observed \sisp\ spectrum extracted from an off-nucleus region of a CALIFA ETG. 
It can be seen that the fit residuals are comparatively small, of the order of 1--2$\sigma$ of the continuum in the blue spectral range, yet systematic.
\rem{lower panel:} The same spectrum after rectification of the rest-continuum.}
\label{fig:SL_residuals}
\end{figure*}

To alleviate this problem, \p3d\ performs for each \sisp\ a rectification
of the rest-continuum (i.e. $O_{\lambda}$--$M_{\lambda}$), which is based
on an 1D adaptation of the flux-conserving unsharp masking technique by
\cite{P98}.  This approach permits determination of a smooth version for
the rest-continuum which is then subtracted from the input data array,
yielding a \rem{net} emission-line spectrum with a mean value of zero
unless emission lines are present (cf lower panel of
Fig.~\ref{fig:SL_residuals}). Other than the more conventional approach of
re-fitting the residual continuum with a high-order polynomial, this
technique captures and corrects both for large and small-scale fitting
residuals, while leaving noise statistics unaffected.

The determination of emission-line fluxes from rectified \rem{net} spectra
is based on the combined application of two methods. The first one
(\rem{el1}) employs standard Gaussian line fitting/deblending with the
Levenberg-Marquardt non-linear fitting algorithm and the second one
(\rem{el2}) a simple summation of the flux within pre-defined spectral
windows centered on the laboratory wavelength of up to 50 emission lines,
after correction for local motions.  In the first case, the previous
rectification of the residual continuum to a zero level is a key advantage,
facilitating fast convergence and enhancing the robustness of emission line
measurements, thereby permitting to extend the analysis down to
EWs$\sim$0.5~\AA\ for a typical S/N$\ga$50 in the continuum. This is
largely because the continuum rectification effectively eliminates two free
parameters (level and slope of the local continuum) in fitting/deblending.
In cases where the deblending of the \ha\ line from the [N{\sc ii}] lines
has failed to converge into a sensible solution, e.g., the [N{\sc
    ii}]$\lambda$6584/\ha\ ratio deviates from $\sim$3 beyond certain
tolerance bounds, \rem{el1} switches to a coarse determination of the
\ha\ flux by fitting the [N{\sc ii}]$\lambda$6584 line only, and assuming a
          [N{\sc ii}]$\lambda$6584/[N{\sc ii}]$\lambda$6548=3.  Prior to
          applying meth. \rem{el2}, \p3d\ uses the gas kinematics pattern
          inferred from \rem{el1} to slightly adjust the center of the
          spectral windows [$\lambda_{\rm min}$:$\lambda_{\rm max}$] within
          which the flux summation is performed.

\p3d\ (v.2) estimates uncertainties in line fluxes by consideration in
quadrature of four contributions to the error budget.
The first one - taken into account only for \rem{el1} measurements - reflects 
formal errors in the fit, as obtained from the covariance matrix and is in most cases negligible. 
The second term considers photon statistics following a widely used formula \citep[e.g.,][see also K12]{GD95,cas02}
\begin{equation}  
\sigma = \sigma_{\rm cont} N^{1/2}\left(1 + \frac{\rm EW}{N\Delta\lambda}\right)^{1/2} 
\end{equation} 
where $\sigma_{\rm cont}$ is the uncertainty in the determination of the
local continuum (defined by its noise level), $N$ the width of the region
used to measure the line in pixels, $\Delta\lambda$ the spectral dispersion
in \AA~pix$^{-1}$, and EW the equivalent width of the line in \AA.
Emission-line EWs were computed from the ratio between the line flux
(measured on the \rem{net} emission line spectra) and the rectified model
SED ($M_{\lambda}$).  The third term, which is included in \rem{el2}
measurements only, uses the noise distribution array in connection with an
empirical scaling factor to gauge how the line flux registered within the
spectral window considered could be affected by positive and negative
photon noise peaks at the 1$\sigma$ level.  Note that none of the above
terms captures uncertainties related to the quality of the stellar fitting.
An analytical estimate for the latter is obviously a non-trivial task,
given that it has to also include a realistic description of how errors in
the best-fitting stellar velocity dispersion and stellar metallicity may
affect fluxes of emission lines being embedded within broader stellar
absorption features.  As a remedy to this problem, \p3d\ includes for
\rem{el2} measurements an error contribution that is linked to the local
deviation $\delta_{\rm fit}$ of the fit (i.e. the value of the
rectification array within the respective spectral window
$\delta_{\lambda}$) as $0.68 \cdot \delta_{\lambda} \cdot \mid\!\delta_{\rm
  fit}\!\mid$. This term, which typically dominates for faint
emission-line fluxes, admittedly simplistic, is meant as a minimum attempt
to quantify the cumulative impact of systematic errors in spectral fits on
the recovered emission-line fluxes.  Related to this issue is likely also
the quality of the SSP models adopted.  As we pointed out in K12, the
fluxes of faint (\ewha$\la$1~\AA) emission lines in ETGs may change to up
to 10\% depending on whether BC03 or MILES SSPs are used.  For this reason,
we have taken in this and our previous studies the computationally
expensive approach of performing the full IFS data processing (spectral
fitting and determination of emission lines) twice, both with the BC03 and
MILES SSP library, and to subsequently combine the results.  This was done
with a decision-tree routine which prioritizes the up to four
determinations for each spaxel (\rem{el1} and \rem{el2} for two SSP
libraries) according to their soundness and estimated uncertainties. This
procedure uses an initial sanity check of several quantities (errors, FWHM
of emission lines, and in the case of the \ha+[N{\sc ii}] blend, the
difference between their central wavelengths) and, in a second pass, the
[N{\sc ii}] 6584/6548 ratios.  Preference was generally given to \rem{el1}
determinations, which, whenever available for both BC- and MILES-based fits
were error-weighted averaged. Whenever only one determination has passed
the quality check, it was adopted for the final map, and the same decision
procedure was followed for \rem{el2} measurements which were typically used
in low-S/N, low-\ewha\ spectra.

\subsection{Module m.3: Computation of the $\tau$ ratio \label{P3D-m3}}
The $\tau$ ratio, defined in K12 as the ratio of the Balmer-line luminosity
expected to arise from pure pAGB photoionization to the observed one,
offers a simple yet efficient means for assessing whether or not hot pAGB
stars are capable of powering the extended \wim\ in ETGs.  After our pilot
analysis of the radial distribution of $\tau$ in two ETGs (K12), we
extended this analysis in P13 to the present sample of 32 ETGs to test the
pAGB photoionization hypothesis.

\p3d\ includes a routine that determines the total \lyc\ photon output for
each \SL\ fit by integrating shortwards of the \lyc\ edge the UV continuum
the SEDs of the individual SSPs (BC) composing the best-fitting population
vector. In this procedure, only SSPs for stellar populations older than
$10^8$ yr are considered in order to compute the total \lyc\ output of the
pAGB component only.  The Balmer-line luminosities expected from it are
computed assuming case~B recombination and standard conditions for the gas
($n_{\rm e}=100$ cm$^{-3}$, T$_{\rm e}$=10$^4$ K).  In ETGs, the typically
large uncertainties in emission line fluxes and their ratios, generally
prevent a reliable determination of the intrinsic extinction in the
\wim. Therefore, the optional inclusion of intrinsic extinction in
\p3d\ \rem{m.3} has to rely on the stellar $V$ band extinction, as inferred
from \SL\ fits, which was generally found to be low ($<$0.1 mag).

As discussed in our previous studies (see also Sect.~\ref{intro}, a $\tau
\simeq 1$ corresponds to an equilibrium situation where the \lyc\ photon
rate from pAGB stars precisely balances the observed \ha\ luminosity.  A
$\tau<1$, on the other hand, indicates an additional contribution by an
extra source, such as AGN, shocks or OB stars. Finally, as we discussed in
P13, a $\tau$ ratio exceeding unity corresponds to the situation of leakage
of a fraction 1-$\tau^{ -1}$ of the \lyc\ output from the pAGB
component. However, given the inherent uncertainties in the currently
available model predictions for the UV output from pAGB stars, we
conservatively assumed in P13 and in this study a $\tau \geq 2$ as robust
evidence for \lyc\ escape.

\subsection{Computation of radial profiles \label{radial_profiles}}
In Sect.~\ref{results} we show the radial distribution of various quantities
of interest, such as the intensity and EW of the \ha\ emission line, and the
\to3hb, \tn2ha\ and $\tau$ ratios. Following our methodology in P13, we
include in all diagrams two complementary data sets:

\begin{figure}[h]
\begin{picture}(8.6,8.6)
\put(0,0){\includegraphics[trim=2cm 2cm 3cm 5cm, clip=true, width=8.00cm,height=8.00cm]{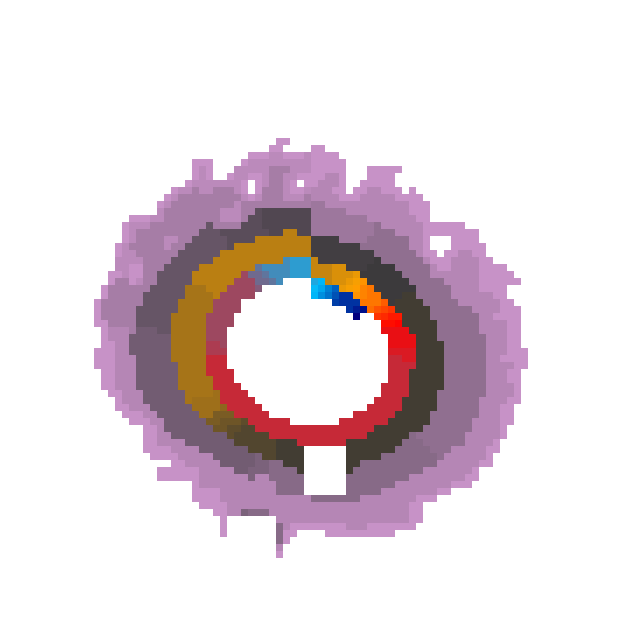}}
\end{picture}
\caption[]{Adaptation of the irregular annuli surface photometry
  technique by \cite{P02} for the purpose of azimuthal binning of spaxels to a
  constant S/N within radial zones. The color coding is meant to illustrate
  the segmentation within isophotal annuli.}
\label{fig:segmentation}
\end{figure}

i) single-spaxel (\sisp) determinations from fits with an 
ADEV $\mid\!\!O_{\lambda}-M_{\lambda}\!\!\mid$ / $O_{\lambda} \leq 2.6$
(cf K12 and the \SL\ manual).  This higher-S/N subsample has the lowest
uncertainties in fitting parameters and emission-line fluxes at the price of
being spatially restricted to typically the central, higher-surface brightness
part ($\mu\la$23 $g$ \sbb) of the studied galaxies.

ii) The average of all \sisp\ determinations within isophotal annuli
(\isan) adapted to the morphology of the (line-free) continuum between 6390
\AA\ and 6490 \AA. These data, which are to be considered in a
statistical sense, go $\ga$2~mag fainter, allowing the study of the
\wism\ in the ETG periphery.

The isophotal annuli method \citep[][referred to as meth.~iv; subsequently
  as {\sl Lazy} by Noeske et al. 2003,2006]{P02} was developed with the
goal of the derivation of radial surface brightness and nebular line
intensity profiles in irregular galaxies with a not well constrained
optical center, for which the conventional approach of fitting ellipses to
isophotes of 2D axis-symmetric luminosity components is of limited
applicability.  Its main concept consists in the determination of photon
statistics within equidistant logarithmic slices of a galaxy image (or of a
reference image, such as, e.g., a stacked image of all available passbands,
or in the context of this study the line-free 6390--6490
\AA\ pseudo-continuum image) with each of these irregular slices faithfully
reproducing the morphology of a galaxy within each intensity interval
[$I$:$I$+$\delta I$] and corresponding to a well-defined photometric radius
\rr.  In the context of IFS data, this technique was first used in K12 (see
their Appendix for further details) and more recently by
\cite{Sanchez2014b} for the derivation of gas-phase metallicity gradients
in CALIFA late-type galaxies.
  
Figure~\ref{fig:segmentation} illustrates the adaptation of the isophotal
annuli technique such as to permit azimuthal binning of spaxels to a
constant S/N within radial zones. This add-on routine to \p3d\ was used in
P13 and in this study in order to double-check the presence of extended
nebular emission in ETGs.  These tests were supplemented through the
modeling of the integrated spectrum within each annulus, extracted by the
same routine by requesting an unrealistically high S/N.
Figure~\ref{fig:NetSpecInZones} shows examples of the residual emission
after subtraction of the stellar fit with BC SSPs
and continuum rectification within successively large annuli for three ETGs
from our sample, illustrating the extreme faintness of nebular emission in
type~ii ETGs (\object{NGC 6338} and \object{NGC 7550}; panels b\&c) as
compared to type~i ETGs (\object{NGC 1167}; panel a).

\begin{figure*}[t]
\begin{picture}(18.0,14)
\put(0.5,0.0){\includegraphics[height=13.0cm]{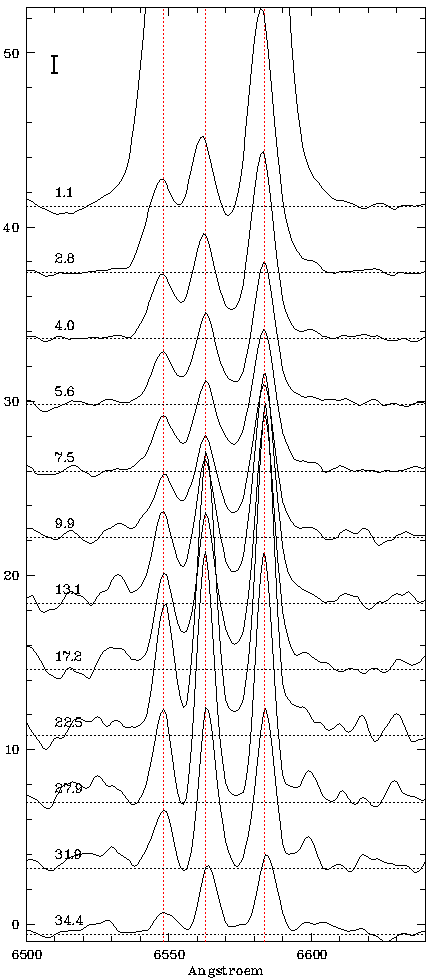}}
\put(6.5,0.0){\includegraphics[height=13.0cm]{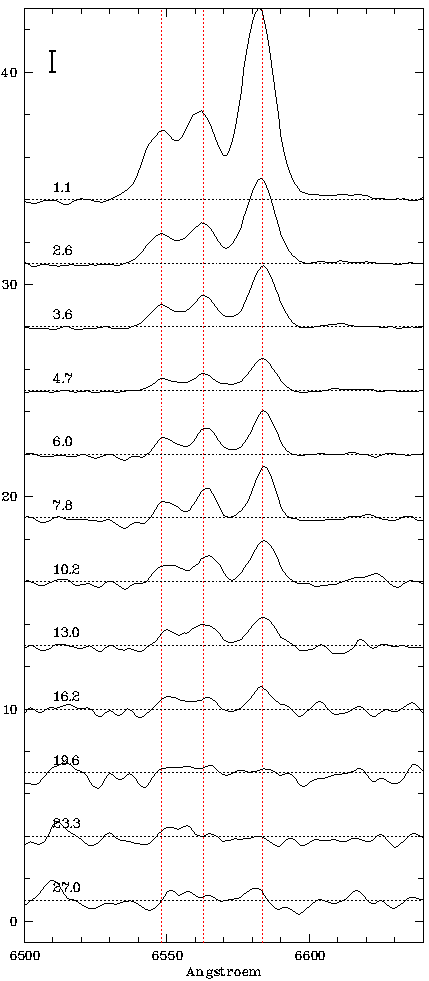}}
\put(12.5,0.0){\includegraphics[height=13.0cm]{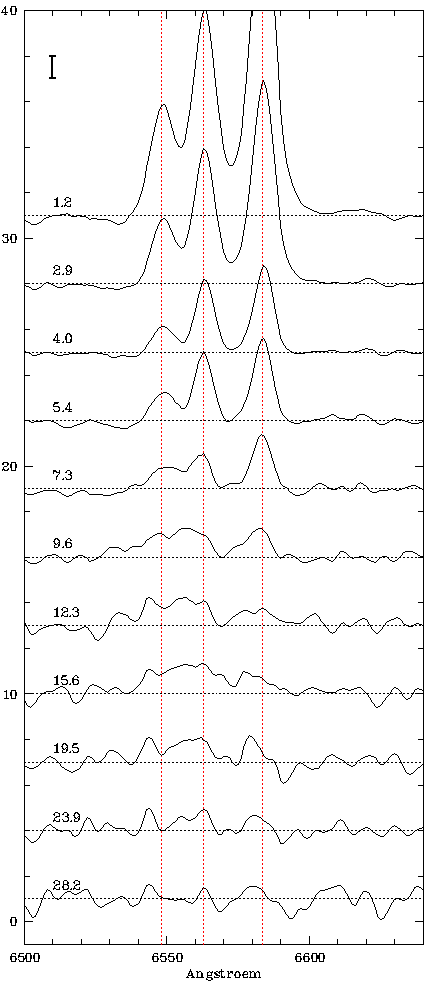}}
\PutLabel{1.2}{12.4}{\vcap a)}
\PutLabel{7.2}{12.4}{\vcap b)}
\PutLabel{13.2}{12.4}{\vcap c)}
\end{picture}
\caption[]{Integrated spectra of the type~i ETG \object{NGC 1167} (\rem{a}) and the type~ii ETGs \object{NGC 6338} and \object{NGC 7550}
(panels \rem{b} and \rem{c}, respectively) within successive isophotal annuli (\isan), after subtraction of the stellar emission 
and rectification of the continuum.
Each spectrum is labeled with the photometric radius (\arcsec) of the annulus within which it was sampled and vertically shifted 
by an arbitrary amount, with the overlaid dotted horizontal lines depicting the zero level. The vertical bar in each panel corresponds to 
a flux level interval of $10^{-16}$ \uflux. The extreme faintness of extended nebular emission in type~ii ETGs, as compared to the type~i 
ETG \object{NGC 1167} is apparent.}
\label{fig:NetSpecInZones}
\end{figure*}


\section{2D emission-line patterns and kinematics for 32
  galaxies \label{ETGsample}}

This section illustrates the diversity of ETG galaxies within our sample with
respect to morphology, kinematics and physical properties of their warm
interstellar medium. The galaxies are ordered according to
Table~\ref{tab:sample}. Note that the maps and radial pro- files
(\rem{a}--\rem{n}) follow the same definitions and colors as in
Sect. \ref{results}.

\plotETG[arqO=UGC_05771,arqT=u05771  ,BPTs=bpt_u05771,name=UGC 5771 ,type=i,morp=S0/a    ,tauR=0.69,dist=106.4,Magr=-21.9 ,lumH=34.0 ,Mass=11.398,desX=0,desY=0] \\
\plotETG[arqO=NGC_4003 ,arqT=ngc4003 ,BPTs=bpt_n4003 ,name=NGC 4003 ,type=i,morp=SB0     ,tauR=0.57,dist=96.6 ,Magr=-21.71,lumH=50.8 ,Mass=11.218,desX=0,desY=0] \\
\plotETG[arqO=UGC_8234 ,arqT=u08234  ,BPTs=bpt=u08234,name=UGC 8234 ,type=i,morp=SO/a    ,tauR=1.09,dist=116.1,Magr=-22.5 ,lumH=15.9 ,Mass=11.131,desX=0,desY=0] \\
\plotETG[arqO=NGC_5966 ,arqT=n5966   ,BPTs=bpt_n5966 ,name=NGC 5966 ,type=i,morp=E       ,tauR=1.26,dist=69.0 ,Magr=-21.67,lumH=117.2,Mass=11.155,desX=0,desY=0] \\
\plotETG[arqO=UGC_10205,arqT=u10205  ,BPTs=bpt_u10205,name=UGC 10205,type=i,morp=Sa      ,tauR=0.20,dist=97.6 ,Magr=-21.79,lumH=63.9 ,Mass=11.221,desX=0,desY=0] \\
\plotETG[arqO=NGC_6081 ,arqT=n6081   ,BPTs=bpt_n6081 ,name=NGC 6081 ,type=i,morp=S0      ,tauR=1.07,dist=79.6 ,Magr=-21.64,lumH=18.9 ,Mass=11.331,desX=0,desY=0] \\
\plotETG[arqO=NGC_6146 ,arqT=n6146   ,BPTs=bpt_n6146 ,name=NGC 6146 ,type=i,morp=E?      ,tauR=1.86,dist=127.4,Magr=-23.04,lumH=34.9 ,Mass=11.720,desX=0,desY=0] \\
\plotETG[arqO=UGC_10695,arqT=u10695  ,BPTs=bpt_u10695,name=UGC 10695,type=i,morp=E       ,tauR=0.90,dist=120.1,Magr=-22.21,lumH=47.3 ,Mass=11.335,desX=0,desY=0] \\
\plotETG[arqO=UGC_10905,arqT=ugc10905,BPTs=bpt_u10905,name=UGC 10905,type=i,morp=S0/a    ,tauR=0.98,dist=114.1,Magr=-22.26,lumH=33.8 ,Mass=11.487,desX=0,desY=0] \\
\plotETG[arqO=NGC_6762 ,arqT=n6762   ,BPTs=bpt_n6762 ,name=NGC 6762 ,type=i,morp=S0/a    ,tauR=0.77,dist=44.9 ,Magr=-20.07,lumH=56.7 ,Mass=11.524,desX=0,desY=0] \\
\plotETG[arqO=NGC_7025 ,arqT=ngc7025 ,BPTs=bpt_n7025 ,name=NGC 7025 ,type=i,morp=Sa      ,tauR=1.67,dist=70.7 ,Magr=-22.05,lumH=36.1 ,Mass=11.496,desX=0,desY=0] \\
 
\plotETG[arqO=NGC_1349 ,arqT=n1349   ,BPTs=bpt_ngc1349,name=NGC 1349,type=i+,morp=S0     ,tauR=0.89,dist=87.7 ,Magr=-21.62,lumH=42.9 ,Mass=11.183,desX=0,desY=0] \\
\plotETG[arqO=NGC_3106 ,arqT=n3106   ,BPTs=bpt_ngc3106,name=NGC 3106,type=i+,morp=S0     ,tauR=0.60,dist=90.1 ,Magr=-22.08,lumH=63.0 ,Mass=11.314,desX=0,desY=0] \\
 
\plotETG[arqO=UGC_0029 ,arqT=ugc0029 ,BPTs=bpt_u0029 ,name=UGC 0029 ,type=ii,morp=E      ,tauR=4.98,dist=118.5,Magr=-21.93,lumH=5.6  ,Mass=11.203,desX=0,desY=0] \\
\plotETG[arqO=NGC_2918 ,arqT=n2918   ,BPTs=bpt_n2918 ,name=NGC 2918 ,type=ii,morp=E      ,tauR=4.87,dist=98.1 ,Magr=-22.40,lumH=6.9  ,Mass=10.340,desX=0,desY=0] \\
\plotETG[arqO=NGC_3300 ,arqT=n3300   ,BPTs=bpt_n3300 ,name=NGC 3300 ,type=ii,morp=SAB(r)0,tauR=19.8,dist=48.0 ,Magr=-20.96,lumH=1.1  ,Mass=10.858,desX=0,desY=0] \\
\plotETG[arqO=NGC_3615 ,arqT=n3615   ,BPTs=bpt_n3615 ,name=NGC 3615 ,type=ii,morp=E      ,tauR=10.8,dist=98.3 ,Magr=-22.46,lumH=4.0  ,Mass=11.523,desX=0,desY=0] \\
\plotETG[arqO=NGC_4816 ,arqT=n4816   ,BPTs=bpt_n4816 ,name=NGC 4816 ,type=ii,morp=S0     ,tauR=18.6,dist=102.6,Magr=-22.15,lumH=4.4  ,Mass=11.367,desX=0,desY=0] \\
\plotETG[arqO=NGC_6125 ,arqT=ngc6125 ,BPTs=bpt_n6125 ,name=NGC 6125 ,type=ii,morp=E      ,tauR=23.4,dist=72.2 ,Magr=-22.16,lumH=3.2  ,Mass=11.437,desX=0,desY=0] \\
\plotETG[arqO=NGC_6150 ,arqT=n6150   ,BPTs=bpt_n6150 ,name=NGC 6150 ,type=ii,morp=E?     ,tauR=10.9,dist=126.0,Magr=-22.38,lumH=3.8  ,Mass=11.496,desX=0,desY=0] \\
\plotETG[arqO=NGC_6173 ,arqT=n6173   ,BPTs=bpt_n6173 ,name=NGC 6173 ,type=ii,morp=E      ,tauR=15.6,dist=126.9,Magr=-23.19,lumH=8.6  ,Mass=11.999,desX=0,desY=0] \\
\plotETG[arqO=UGC_10693,arqT=u10693  ,BPTs=bpt_u10693,name=UGC 10693,type=ii,morp=E      ,tauR=11.7,dist=120.5,Magr=-22.70,lumH=5.9  ,Mass=11.572,desX=0,desY=0] \\
\plotETG[arqO=NGC_6338 ,arqT=n6338   ,BPTs=bpt_n6338 ,name=NGC 6338 ,type=ii,morp=S0     ,tauR=2.21,dist=117.5,Magr=-22.79,lumH=29.4 ,Mass=11.707,desX=0,desY=0] \\
\plotETG[arqO=NGC_6411 ,arqT=n6411   ,BPTs=bpt_n6411 ,name=NGC 6411 ,type=ii,morp=E      ,tauR=8.62,dist=57.6 ,Magr=-21.71,lumH=3.9  ,Mass=11.203,desX=0,desY=0] \\
\plotETG[arqO=NGC_6427 ,arqT=n6427   ,BPTs=bpt_n6427 ,name=NGC 6427 ,type=ii,morp=S0     ,tauR=7.60,dist=51.3 ,Magr=-20.84,lumH=2.1  ,Mass=10.985,desX=0,desY=0] \\
\plotETG[arqO=NGC_6515 ,arqT=n6515   ,BPTs=bpt_n6515 ,name=NGC 6515 ,type=ii,morp=E      ,tauR=2.46,dist=99.0 ,Magr=-22.11,lumH=10.2 ,Mass=11.316,desX=0,desY=0] \\
\plotETG[arqO=NGC_7194 ,arqT=n7194   ,BPTs=bpt_n7194 ,name=NGC 7194 ,type=ii,morp=E      ,tauR=12.9,dist=110.7,Magr=-22.35,lumH=4.4  ,Mass=11.536,desX=0,desY=0] \\
\plotETG[arqO=NGC_7236 ,arqT=n7236   ,BPTs=bpt_n7236 ,name=NGC 7236 ,type=ii,morp=SA0    ,tauR=8.16,dist=108.2,Magr=-21.45,lumH=5.0  ,Mass=11.403,desX=0,desY=0] \\
\plotETG[arqO=UGC_11958,arqT=u11958  ,BPTs=bpt_u11958,name=UGC 11958,type=ii,morp=SA0    ,tauR=2.67,dist=108.3,Magr=-21.82,lumH=21.4 ,Mass=11.612,desX=0,desY=0] \\
\plotETG[arqO=NGC_7436B,arqT=n7436B  ,BPTs=bpt_n7436B,name=NGC 7436B,type=ii,morp=E      ,tauR=6.87,dist=100.6,Magr=-22.21,lumH=11.4 ,Mass=11.708,desX=0,desY=0] \\
\plotETG[arqO=NGC_7550 ,arqT=n7550   ,BPTs=bpt_n7550 ,name=NGC 7550 ,type=ii,morp=SA0    ,tauR=3.02,dist=69.2 ,Magr=-21.91,lumH=15.3 ,Mass=11.394,desX=0,desY=0] \\
 
\section{Comparison with star-forming/composite galaxies \label{LTGsample}}
 
This appendix is meant to illustrate the variation of the \ewha, $\tau$;
and diagnostic line ratios from early-type towards late-type galaxies. All
but one of these systems are classifiable by BPT ratios in their central
part and/or extranuclear component as {\sl composite} or SF-{\sl
  dominated}. \\
 
 \clearpage
 
\plotCom[arqO=IC_0540 ,arqT=ic0540,BPTs=bpt_ic0540,name=IC 540  ,morp=S  ,tauR=0.36,dist=31.9 ,Magr=-18.91,lumH=4.9 ,Mass=10.004] \\
\plotCom[arqO=UGC_8778,arqT=u08778,BPTs=bpt_u8778 ,name=UGC 8778,morp=S? ,tauR=0.31,dist=52.6 ,Magr=-20.04,lumH=19.6,Mass=10.439] \\
\plotCom[arqO=IC_0944 ,arqT=ic0944,BPTs=bpt_ic0944,name=IC 944  ,morp=Sa ,tauR=0.58,dist=104.8,Magr=-22.21,lumH=70  ,Mass=10.439] \\
\plotCom[arqO=IC_1683 ,arqT=ic1683,BPTs=bpt_ic1683,name=IC 1683 ,morp=S? ,tauR=0.18,dist=65.4 ,Magr=-20.61,lumH=87  ,Mass=10.772] \\
\plotCom[arqO=NGC_4470,arqT=n4470 ,BPTs=bpt_n4470 ,name=NGC 4470,morp=Sa?,tauR=0.03,dist=38.4 ,Magr=-20.35,lumH=147 ,Mass=10.108] \\

\end{document}